\documentclass[12pt,preprint]{aastex}

\shorttitle{TRACING THE DARK MATTER STRUCTURE OF Cl~0024+17 WITH ICL AND ICM}
\shortauthors{Jee et al.}

\begin{document}

\title{TRACING THE PECULIAR DARK MATTER STRUCTURE IN THE GALAXY CLUSTER Cl~0024+17 WITH INTRACLUSTER STARS AND GAS} 

\author{M.J. JEE}
	
\begin{abstract}
Intracluster light (ICL) is believed to originate from the stars stripped from cluster
galaxies. They are no longer gravitationally bound to individual galaxies, but
to the cluster, and
their smooth distribution potentially makes them serve as much denser tracers of the cluster
dark matter than the sparsely distributed cluster galaxies. 
We present our study of the ICL in the galaxy cluster Cl~0024+17 using both
Advanced Camera for Surveys (ACS)
and Subaru data, where we previously 
reported discovery of a ring-like dark matter structure with gravitational
lensing. The ACS images provide much lower sky levels than ground-based
data, and enable us to measure relative variation of surface brightness reliably.
This analysis is
repeated with the Subaru images to examine if consistent features are recovered despite different 
reduction scheme and instrumental characteristics.
We find that the intracluster light profile clearly resembles the peculiar mass profile, which
stops decreasing at $r\sim50\arcsec$ ($\sim265$ kpc) and slowly increases until it turns over 
at $r\sim75\arcsec$ ($\sim397$ kpc).
This feature is seen in both ACS and Subaru images for nearly all available passband images
while the features are in general stronger in red filters.
The consistency across different filters and instruments strongly rules out the possibility that 
the feature might come from any residual, uncorrected 
calibration errors.
In addition, our re-analysis of the cluster X-ray data shows that the peculiar mass structure is also indicated
by a non-negligible (3.7~$\sigma$ in $Chandra$ and 2.4~$\sigma$ in XMM-$Newton$) bump in the intracluster gas profile 
when the geometric center of the dark matter ring, {\it not the peak 
of the X-ray emission},
is chosen as the center of the radial bin. 
The location of the gas ring is closer to the center
by $\sim15\arcsec$ ($\sim80$ kpc), raising an interesting possibility that the ring-like structure is expanding and
the gas ring is lagging behind perhaps because of the ram pressure {\it if} both features in mass and
gas share the same dynamical origin.
\end{abstract}

\altaffiltext{1}{Department of Physics, University of California, Davis, One Shields Avenue,
Davis, CA 95616}

\keywords{gravitational lensing ---
dark matter ---
cosmology: observations ---
X-rays: galaxies: clusters ---
galaxies: clusters: individual (\objectname{Cl~0024+17}) }

\section{INTRODUCTION \label{section_introduction}}
Intracluster light (ICL), which was discovered by Zwicky (1951), is believed to originate from the stars stripped 
from cluster galaxies.
Although details are still in dispute, a number of studies suggest
that both the scatter of stars during the brightest cluster galaxy formation (e.g, Gerhard et al. 2007) and 
the tidal disruption of dwarf galaxies (e.g., Mihos et al. 2005) are among the dominant
mechanisms of the ICL production.
These stars are in general considered bound not to any
individual galaxy, but to the cluster mass. 
If a large fraction of the ICL is produced during the assembly of the
brightest cluster galaxies (e.g., Murante et al. 2004), the dynamically collisionless
property of the intracluster stars potentially allows us to use ICL as visible tracers of
underlying dark matter at least in the central region of the cluster.

Observationally, however, the quantitative study of ICL is difficult. 
The typical surface
brightness of ICL is often quoted as $\sim1$\% or less of the night sky from the ground.
At this faint level, instrumental
systematic effects such as residual flat-fielding, scattered lights
of bright stars, etc. become critical issues. In addition, most galaxy edges continuously blend
into ICL, which obviously causes an ambiguity in determining where the galaxy light stops and ICL begins.

In this paper, we present a study of ICL in the galaxy cluster Cl 0024+17 at $z=0.4$, where we recently
discovered a peculiar ringlike dark matter structure (Jee et al. 2007; J07 hereafter). The two-dimensional mass 
reconstruction of J07 shows that the core of the cluster is surrounded by a $r\sim0.4$ Mpc ring-like overdense region.
The feature is strongly constrained by 
coherent fluctuation of background galaxy shapes across the $r\sim0.4$ Mpc circle (weak-lensing) signaling the sudden change of the density slope.
Of course, it is tempting to try to detect the peculiar dark matter structure with 
the cluster galaxies (e.g., Qin et al. 2008). However, because the density contrast of the feature with respect to the 
neighboring region is low ($<5$\%), the sparsely distributed galaxies cannot provide sufficient statistics to 
overcome the shot noise even if the cluster galaxy membership identification is next to perfect.
This is the reason that here we investigate, instead, the intracluster stars, which
diffusely distribute in the cluster potential and thus sample the underlying dark matter much more densely.
In this investigation, the measurement accuracy is not limited by the poissonian scatter as
in the study using galaxies, but by the ability to control various systematics critical to precision surface photometry.

We employ two sets of data: {\it Hubble Space Telescope}  Advanced Camera for Surveys (HST/ACS) and Subaru/Suprime Cam images. 
Space-based imaging provides significant advantage over ground-based effort mainly because the sky
is substantially darker (e.g., $\sim1$ and $\sim3$ magnitude fainter in $r$ and $z$ band, respectively).
Also, as the instrument is above the atmosphere, there is no cumbersome, time-dependent airglow effect, which
often prevents stable calibration. However, unfortunately, there have been no ICL studies so far with the HST/ACS data. This
is because the pipeline flatfielding accuracy $\sim1$\% (Mack et al. 2002) has been considered insufficient to 
enable such studies. Therefore, we undertake this time-consuming task of producing/verifying HST/WFC
flats using ``blank images'' before we proceed with the ICL measurements. One drawback in using ACS data is
the small field of view of the instrument, which only covers a $3\arcmin\times3\arcmin$ region. Because the existing
ACS data of the cluster were taken without any plan for this kind of study, the instrument pointing was almost
fixed to the center of the cluster (with only a few tens of pixels dither to fill the two CCD gap), and thus
a care must be taken to minimize the ICL contamination in determining the background level.

Although the sky is much brighter, the Subaru/Suprime Cam images provide a few critical crosschecks. First, the large field
of view of the Suprime Cam allows us to estimate the background level directly, which is important in determining
the net amount of ICL and assessing the degree of the aforementioned ICL contamination in the HST/ACS measurements.
Second, the Subaru data include the NB$_{912}$ narrow band image, which Kodama et al. (2004) used
to probe H$\alpha$ emission as a measure of on-going star formation of the cluster at $z=0.4$. The sky level in this narrow band
is about 1.5 mag darker than in the broadband $z^{\prime}$ filter while the narrow passband
gives high contrast to the H$\alpha$ emission at $z=0.4$ relative to continuum sources at different redshifts.
Third, various instrumental signatures including the point spread function (PSF), internal reflection, obscuration, 
geometric distortion, etc. are different from ACS. Hence, if a significant feature clearly revealed
in one instrument, is absent in the other, this indicates that the feature might come from
some residual calibration errors or time-dependent sources. 
On the contrary, detection of a consistent feature in both images provides evidence against these systematics.

The hot intracluster gas also samples the underlying dark matter very densely although its collisional nature
biases the distribution from that of dark matter. 
A common isothermal $\beta$ model assumes $\rho_{gas} \propto \rho_{DM}^{\beta}$ and $\epsilon_{ff} \propto n_e^2$, where
$\epsilon_{ff}$, and $n_e$ are the free-free radiation emissivity and the electron density, respectively. Therefore, X-ray profiles
in general are more centrally peaked than mass profiles.
Also, as observed
in many merging systems (Jee et al. 2005; Clowe et al. 2006), some gas peaks are often significantly displaced with respect
to dark matter peaks, which is interpreted as signaling the different physical property 
(e.g., collisional versus collisionless nature).
Therefore, a rigorous simulation is desired to fully address whether or not it is feasible to detect the analogous ring-like
structure in the gas profile of Cl~0024+17, given the low level of the contrast in the dark matter structure.
In this paper, we approach the issue purely observationally by
re-analyzing the existing $Chandra$ and XMM-$Newton$ X-ray data.
The current analysis is different
from J07 only in that now we choose the geometric center of the dark matter ring as our origin in setting up radial
bins. If there is indeed any feature in the gas profile hinting at the peculiar dark matter substructure, 
the signal is expected to be of low contrast and thus should be sensitive to the choice of the origin.
Therefore, it is quite possible that J07 might have failed to detect the feature in X-ray because the center was
not placed at the optimal location.

In this paper, we adopt a $\Omega_M=0.3$, $\Omega_{\Lambda}=0.3$, and $h=0.7$ cosmology, where
the plate scale is 5.3 kpc$/\arcsec$ at the redshift of the cluster $z=0.4$.
The quoted uncertainties are at the $1-\sigma$ ($\sim68$\%) level.

\section{OBSERVATIONS AND REDUCTIONS \label{section_obs}}

\subsection{ACS Data Reduction}
The cluster was observed with the Wide Field Channel (WFC) of the
ACS in 2004 November. A single pointing ($\sim3\arcmin.3 \times \sim3\arcmin.3$ field of view) is centered
near the cluster core ($\alpha_{2000}\simeq00^h:26^m:35^s, \delta_{2000}\simeq17\degr:09\arcmin:43\arcsec$) with
integrations of 6435~s, 5072~s, 5072~s, 8971~s, 10144~s, and 16328~s in the F435W, F475W, F555W, F625W,
F775W, and F850LP filters, respectively.
The low level CCD processing was carried out using the STScI standard ACS calibration pipeline (CALACS; Hack et al. 2003)
with some important modifications as follows.
CALACS applies the LP-flats derived from
both the laboratory before launch (Bohlin et al. 2001) and the 47 Tuc stars on
orbit (Mack et al. 2002). We find that these pipeline flats deviate from sky
flats at the $\lesssim$0.5\% level (see \textsection\ref{section_acs_flat} for
details). This level of inaccuracy should be corrected prior to ICL measurement 
although the sky level of HST images is 1-3 mags fainter.
We applied these residual sky flats to these CALACS-processed FLT images. Careful inspection of these results reveal that
the amplifier-dependent (quadrant-to-quadrant variation) bias at the level of $\sim1$ DN were present in some frames. 
We fix these quadrant-to-quadrant variation by comparing the medians of the two adjacent strips ($\sim100$ pixel width) evaluated
after masking out astronomical objects and cosmic rays. In principle, these additive biases are supposed to be corrected 
before the application of the flats, which are multiplicative. However, as the bias is small and the flats are already close to unity,
the errors arising from reversing the order of these two operations are negligible. 

We measured sky levels in each frame through iterative sigma-clipping and subtracted
the values before stacking. Table 1 summarizes mean sky levels for each filter. 
The final high-level processing involving geometric distortion correction and cosmic-ray removal
was performed using the ``apsis" pipeline (Blakeslee et al. 2003).
We used nearest neighbor interpolation in drizzling (Fruchter and Hook 2002). This is different from the procedure of J07,
where the Lanczos3 (windowed sinc function) kernel is employed. The Lanczos3 kernel provides a sharp point spread function, and
thus suitable for weak-lensing studies. However, this kernel correlates neighboring pixels and artificially reduces noise
fluctuation ($\sim8$\%) although the degree of the correlation is found to be much lower than other commonly selected
kernels such as square, gaussian, etc (Mei et al. 2005). Our choice of the nearest neighbor interpolation is to
prevent any potential distortion of the intrinsic noise power spectrum due to drizzling.

\subsection{Subaru Data Reduction \label{section_subaru_reduction}}

We retrieved the archival Subaru images of the cluster from SMOKA\footnote{http://smoka.nao.ac.jp/}.
The data were taken in 2002 September 7 with the prime-focus camera (Suprime-Cam; Miyazaki et al. 2002) 
in the NB$_{912}$ ($\lambda_{eff}$=9139~\AA, FWHM=134~\AA), $B$, $R_c$, and $z^{\prime}$ 
filters with integrations of 10,800~s, 3,600~s, 5,280~s, and 1,980~s, respectively. The number of visits per filter
and the sky level variation are summarized in Table 2. The Suprime Cam focal plane is tiled with 2$\times$5 CCDs 
covering an area of $30\arcmin\times27\arcmin$. Because we are only interested in the central region of the cluster, 
we only used the central six CCDs, which still cover a significantly large
area ($20\arcmin \times 27\arcmin$).

The Suprime-Cam data reduction software SDFRED (Yagi et al. 2002; Ouchi et al. 2004) was used 
to subtract overscan, mask out regions affected by Acquisition and Guide (AG) probe and bad pixels, 
perform flatfielding (see \ref{section_subaru_flat} for details of our
sky flat creation), and correct geometric distortion.
The sky subtraction procedure of SDFRED removes small-scale sky variations, which
undesirably distorts the ICL structure and the large scale PSF profiles.
Thus, we replace this step with our manual procedure summarized as follows.
Each flatfielded, $20\arcmin\times27\arcmin$ frame shows $\lesssim1$\% sky gradient. This type of
residual gradient is common and is attributed to night-by-night atmospheric effects
rather than instrumental flat changes. In general, the pattern is monotonic and can be
removed by modeling the gradient with a low-order polynomial plane. In this procedure,
a care must be taken to mask 
out extended PSF wings of bright stars, galactic cirrus, and the central region of the cluster ($3\arcmin\times3\arcmin$), as well
as resolved astronomical objects.
Masking out PSF wings of bright stars was not difficult
for several $m_r=10\sim12$ mag stars inside the image. 
However, there is one $m_r\lesssim7.5$ mag star
outside ($\sim 1 \arcmin$ away) the camera's field of view, which causes severe 
scattered light in the southwestern corner. 
Because of the large scale ($\sim4\arcmin$) of the feature, it is difficult to disentangle
this scattered diffuse light from the sky gradient.
Therefore, we chose to model/subtract
this scattered light and the residual sky gradient at the same time by fitting a 4$^{th}$ order polynomial plane
to the entire
field. It is possible that any imperfection in this procedure increases a large-scale sky level fluctuation beyond
the flatfielding accuracy. Therefore, we examined sky levels at random locations, and verified that
the variation is consistent with what we expect from the flatfielding errors.

After we subtracted the mode from each frame, we determined the spatial alignment and flux scaling between frames
using bright non-saturated stars. This information was provided to SDFRED to create the final mosaic images.
Zeropoints were evaluated by comparing photometry with that measured from the ACS images. We verified that these zeropoints are
consistent with the values independently determined by Kodama et al. (2004), who kindly provided us with their results\footnote{Kodama 
et al. (2004) used Vega-based magnitude system while in the current paper all the magnitudes are given in the AB system}.

We display in the top panel of Figure~\ref{fig_ICL_2d} the color-composite image of the cluster from the Subaru mosaic images.
The image shows the central $10\arcmin\times10\arcmin$ field with the blue, green, and red intensities representing
the $B$, $R_c$, and $z^{\prime}$ fluxes. White circles denote the spectroscopic cluster members ($0.37<z<0.42$) that we
obtained from the publicly available Moran et al. (2005) catalog.
The yellow line delineates the $3\arcmin\times3\arcmin$ ACS field, which is also separately shown
on the lower left corner. The lower right panel displaces the diffuse light of the cluster measured from
the ACS F625W image after objects are subtracted/masked out(see \ref{section_masking}).

\subsection{Flat-Fielding}

\subsubsection{ACS Flat-fielding \label{section_acs_flat}}

The ACS flats used by the current pipeline of the STScI, called LP-flats, were
derived by combining the factory-measured P-flats before launch (Bohlin et al. 2001) and the inflight L-flats (Mack et al. 2002), 
which correct the pixel-to-pixel and the low-frequency variations, respectively. The L-flats were created directly
for the filters F435W, F555W, F606W, F775W, F814W, and F850LP by observing stars in 47 Tuc, and
indirectly for the remaining filters by interpolating the direct measurement results. 
Mack et al. (2005) report that the LP-flats are expected to reduce field-dependent sensitivity variation down to $\lesssim$1\% 
for the first six filters, and $\lesssim$2-3\% for the interpolated filters.
Pavlovsky et al. (2006) mention that their
investigation of the sky flats constructed from GOODS images for the F606W, F775W, and F850LP filters are in good ($<2$\%) agreement
with the pipeline LP-flats.

The $\sim2$\% flatfielding error (i.e., the reported difference between the LP-flats and the STScI sky flats), 
if present on a large scale, is of concern for the investigation of the ICL features that this paper
studies; for example, this will limit our ability to perform surface photometry to the $\mu \lesssim 27\mbox{mag}~\mbox{arcsec}^{-2}$
regime in the F625W filter. Because the sky flats of the STScI are not yet publicly available, it is impossible to determine the
pattern of the deviation and thus the impact of the reported inaccuracy on our ICL measurement.
Therefore, we decided to independently create sky flats for all six ACS filters used for the Cl~0024+17 observations.

We collected $90\sim300$ blank sky images for each filter from the STScI archive.
We avoided images, which have large extended sources (e.g., nearby galaxies), many bright stars (e.g., globular
clusters), or diffuse light (e.g., ICL in galaxy clusters, galactic cirrus, etc.). The selected images were manually examined, and discarded if
any noticeable sky gradient is present. It is well known that quadrant-to-quadrant variation ($1-3$\% of background counts) is 
present in WFC images (see Sirianni et al. 2002 for description of the problem). Unless corrected for, these features are clearly
visible in the final sky flat because the pattern is not completely random. We fix these quadrant-to-quadrant variation
by comparing the medians of the two adjacent strips ($\sim100$ pixel width) after masking out astronomical objects and cosmic
rays. Then, we median-smooth each image with a box size of 32 pixels, and normalize the image by dividing it by the
mode of the image. The final, median-stacked image of these 
normalized, median-smoothed images
shows the residual flatfielding error because we perform the task on FLT files that have already been applied the pipeline LP-flats.
We show in Figure~\ref{fig_flat_all} these residual sky flats\footnote{The FITS images of these ACS residual flats are
publicly available on request}.
Table 3 lists the number of images that we used for the creation of sky flats for each filter, the deviation from
the pipeline flat, and the 
accuracy of the residual sky flat that we estimate by bootstrapping.
The $\lesssim0.1$\% accuracy in flatfielding implies that we can probe the surface brightness limit
down to the $\mu\sim30 \mbox{mag}~\mbox{arcsec}^{-2}$ level in nearly all ACS filters if flatfielding is the only dominant
source of errors.

Our sky flats confirm the claim of Mack et al. (2005) and Pavlosky et al. (2006) on 
the $\lesssim2$\% accuracy of the pipeline
LP-flats. The most notable large scale structure is the donut-like pattern particularly clear in red filters (the
difference in residual between the center of the donut pattern and the trough ($\sim1\arcmin$ away) is 
$\sim0.5$\% and $\sim2$\% in F775W and F850LP, respectively).
This residual feature is also mentioned by Pavlovsky et al. (2006), and they
suggest that this might be due to the
difference in spectrum between 47 Tuc stars and the sky. 
While we agree that the difference in color might be a plausible source of the
residual, we expect the residual to be still observable even if there is no color
difference between 47 Tuc stars and the sky.  The sky flats measure both
sensitivity and projected pixel area (i.e., due to geometric distortion) effects
whereas the latter is difficult to measure without bias using stellar photometry alone 
unless the PSF is sufficiently oversampled.\footnote{Imagine an extreme case
where the PSF is severely undersampled and the stellar profile is confined to
a single pixel. Then, the total flux in each CCD pixel only informs us of
the sensitivity, not the projected pixel size.}

On the center of the donut-like feature in F435W and F475W, there is a strong indication that
the pipeline flats over-correct the sensitivity in this region. Comparing these residual sky flats
with the pipeline flats also reveals that there are a few dust-moats (e.g., bottom of F555W and center of F625W) that were not
included in the pipeline LP-flats. Finally, it is clear that a gridlike pattern 
exists in all six filters shown here. The feature is most (least) obvious in the F850LP (F625W) filter.
The average distance between grids is $\sim60$ pixels. Because this checkerboard pattern does not change in size
and location as we vary smoothing kernels and sizes, we conclude that the pattern is not an artifact of our 32-pixel
median smoothing. The exact cause of the checkerboard pattern has not been known yet. Nevertheless, we suspect that this
particular residual pattern might be the remnant of the ACS CCD fabrication process that is not corrected by the P-flat.
The peak-to-valley variation of the feature amounts to $\sim1$\% in F850LP.

\subsubsection{Subaru Flat-fielding \label{section_subaru_flat}}

We retrieved blank images taken during the 2002 August-September period from the Subaru archive.
As in the case for ACS, we manually examined and discarded frames that are not adequate 
for flatfield generation (e.g., presence of diffuse emission, extremely saturated stars, crowded
stellar fields, etc.).
Bad pixels and the vignetted areas by the Acquisition and Guide (AG) Probe on the top five CCDs were masked out 
in the remaining frames.
Astronomical objects were identified and masked out by searching for 5 or greater continuous
pixels above 1.5 times the sky rms. After median-smoothing each frame with a box size of $16\times16$ pixels, we normalized 
the image using its mode, and then median-stacked all frames to create preliminary sky flats.
These preliminary sky flats allow us to refine the evaluation of the previous modes and the detection
of astronomical objects to be masked out.
Therefore, we obtained the final sky flats by iteratively evaluating the modes, detecting objects, and creating flats
as done by Morrison et al. (1997). One important modification to the Morrison et al. (1997) method
is that we removed low-frequency residual sky gradients ($<1$\%) by fitting
second-order polynomial planes. As noted by many authors
(e.g.,Feldmeier et al. 2002), each frame possesses non-negligible sky gradient due to
many atmospheric effects even if a perfect sky flat is applied. 
This removal of the residual
sky gradient further reduces the width of the sky distribution within a frame, which also
helps us to mask out faint astronomical objects more efficiently.
After a few iterations, we obtained
converged master sky flats. 

The accuracy of these final sky flats were not limited by photon statistics thanks to the large light-collecting
power of the Subaru telescope, but by the aforementioned large-scale sky variations in input frames.
Therefore, we estimated the accuracy of the sky flat by bootstrap-resampling the input frames. 
The estimated error for the six central CCDs that we keep for the cluster reduction is
on average $\sim0.07$\% (see Table 4 for the individual filters).
The $0.07$\% accuracy in flatfielding, if this is the sole source of systematics, allows us to probe the 
surface brightness $\sim8$ mag deeper than the sky level. For the $B$ filter, this surface
brightness limit is $\sim30.5~\mbox{mag}~\mbox{arcsec}^{-2}$, similar to the value in ACS data whereas 
for the $z^{\prime}$ filter, this limit implies that we can reach down to the $\mu\sim27.5 \mbox{mag}~\mbox{arcsec}^{-2}$
level because of the high sky level ($\mu\simeq19.2$ mag~$\mbox{arcsec}^{-2}$). 
However, it is important to remember that the quoted accuracy represents the large-scale 
error within the entire field ($20\arcmin\times27\arcmin$). While this large-scale error
is a major source of uncertainty in our determination of the background level (thus limiting
the accuracy in absolute ICL level measurement), the
flatfielding error affecting the relative significance of the ICL profile is
the uncertainty of the flats within the central $3\arcmin\times3\arcmin$ region.
As this area occupies only $\sim1.7$\% of the total $20\arcmin\times27\arcmin$area,
the flat accuracy within the region is significantly (a factor of two) better than 
the accuracy across the field.
In addition, the dithering of the observation improves the flatness further, turning systematics into
statistics. 
Because the current paper is focused
on the structure of the ICL profile within this region, we distinguish these relative errors
from the absolute errors when necessary.

For a sanity check, we compare these sky flats with dome flats.
For the $B$, $z^{\prime}$, and NB$_{912}$ filters, high S/N dome flats were taken along with the cluster Cl~0024+17 on the same
night (2002 September 7). For the $R_c$ filter, the closest dome flats in time were taken on 2002 September 4.
Again, we limited our analysis to the central six CCDs.
We found an rms difference of $\sim0.2$\% on average, which
suggests that the Suprime-Cam flats are time-stable at least over this two-month period. 
Furthermore, we note that this $\sim0.2$\% discrepancy is dominated by a large scale gradient, and we suspect
that this is caused by the non-uniformity of the dome screen. When this gradient is removed by second-order
polynomial modeling, the agreement becomes $\lesssim0.1$\%.

\subsection{Object Detection and Masking \label{section_masking}}

As it is impossible to model the surface brightness profile of individual galaxies
accurately and to subtract the contribution without introducing biases in our
ICL study, we choose to mask out galaxies. In order to obtain consistent
masking regions across filters, we used a single detection image for each instrument.
For the ACS data set, this image is automatically generated by apsis from weight-averaging
all six filter images. For the Subaru data, the $R_c$-band image is significantly deeper than the rest, and
thus this $R_c$ image is chosen to detect astronomical objects.
We detect objects via SExtractor by searching for at least 5 connected
pixels above 1.5 times sky fluctuation.

Because galaxy light continuously blends into background light, it is important to define
the size of the masking area very carefully. An ideal masking size is to minimize the impact of the
diffuse wing of objects while still leaving a sufficient number of
background pixels usable for the ICL measurement. Obviously, an ellipse defined by
SExtractor's semi-major and minor axes does not sufficiently mask out diffuse wings of galaxy light. Hence, we enlarged
the masking size by integer multiples of these axes (we refer to these as ``$1~\times$", ``$2~\times$", ``$3~\times$", etc.), and examined where our ICL profile starts to converge. 
We observe that the axes of the masking ellipse should be at least three times the value given by SExtractor.
This empirical finding is valid for both ACS and Subaru images.
In Figure~\ref{fig_masking}, we illustrate how much our determination of the location of the
Gaussian peak, which we use in this paper as an indicator of the ICL level, depends on the masking size. In this example, 
the (skewed) Gaussian curves are the histograms of
the pixel values in the F775W image at $r=76\arcsec-84\arcsec$. With insufficient masking, the Gaussian curves
are skewed because the object light spilt outside masking apertures contributes to the bright ends. The suitable
masking size can be determined by examining either the skewness of the Gaussian or the shift of the
centroid between subsequent masking schemes. In this study we adopt the latter as our guiding tool and
find that a subsequent change in the centroid (dashed line) is at the $\mu\gtrsim30~\mbox{mag}~\mbox{arcsec}^{-2}$ level
if the axes of the masking ellipse are three times or larger than the semi-major and -minor axes given by SEXtractor.
When we choose the segmentation 
map (isophotal area) produced by SExtractor (with the 1.5 $\sigma$ detection threshold) instead, the result lies 
between the 3 $\times$ and 4 $\times$ masking schemes (brown).
We note that alternatively some authors set a very low threshold to detect astronomical objects
and use the resulting segmentation map to define the masking areas. We report that
lowering the threshold from 1.5 $\sigma$ to 1.0 $\sigma$ would produce a result close to our ``5 $\times$"
masking case (blue), which we conservatively choose in this paper.

As for stars, we masked only the cores of the stars and subtracted the PSF wings
from the image. We used saturated stars to sample the radial profile of the PSF wing for both ACS and Subaru.
In the STScI archive, we found the saturated images of the star HD39060 (Beta Pictoris, PI:Paul Kalas) for
the F435W, F606W, and F814W filters. The star is located at the center of WFC, and this allows us to
obtain the PSF profile up to $\sim150\arcsec$. The filter-to-filter variation is up to $\sim 1~\mbox{mag}~\mbox{arcsec}^{-2}$
at $r=100\arcsec$ when the profile is normalized using the surface brightness at small radii. We use extrapolation to model the F850LP PSF wing and
interpolation to model the profile for the rest F475W, F555W, F625W, and F775W filters.
For the Subaru/Suprime-Cam data, we used the brightest ($m_R\sim8.6$)
star in the Cl~0024+17 image $\sim7\arcmin$ away from the cluster center. We display these ACS and Subaru PSF
profiles in Figure~\ref{fig_psf_wing}.

\begin{figure}
\plotone{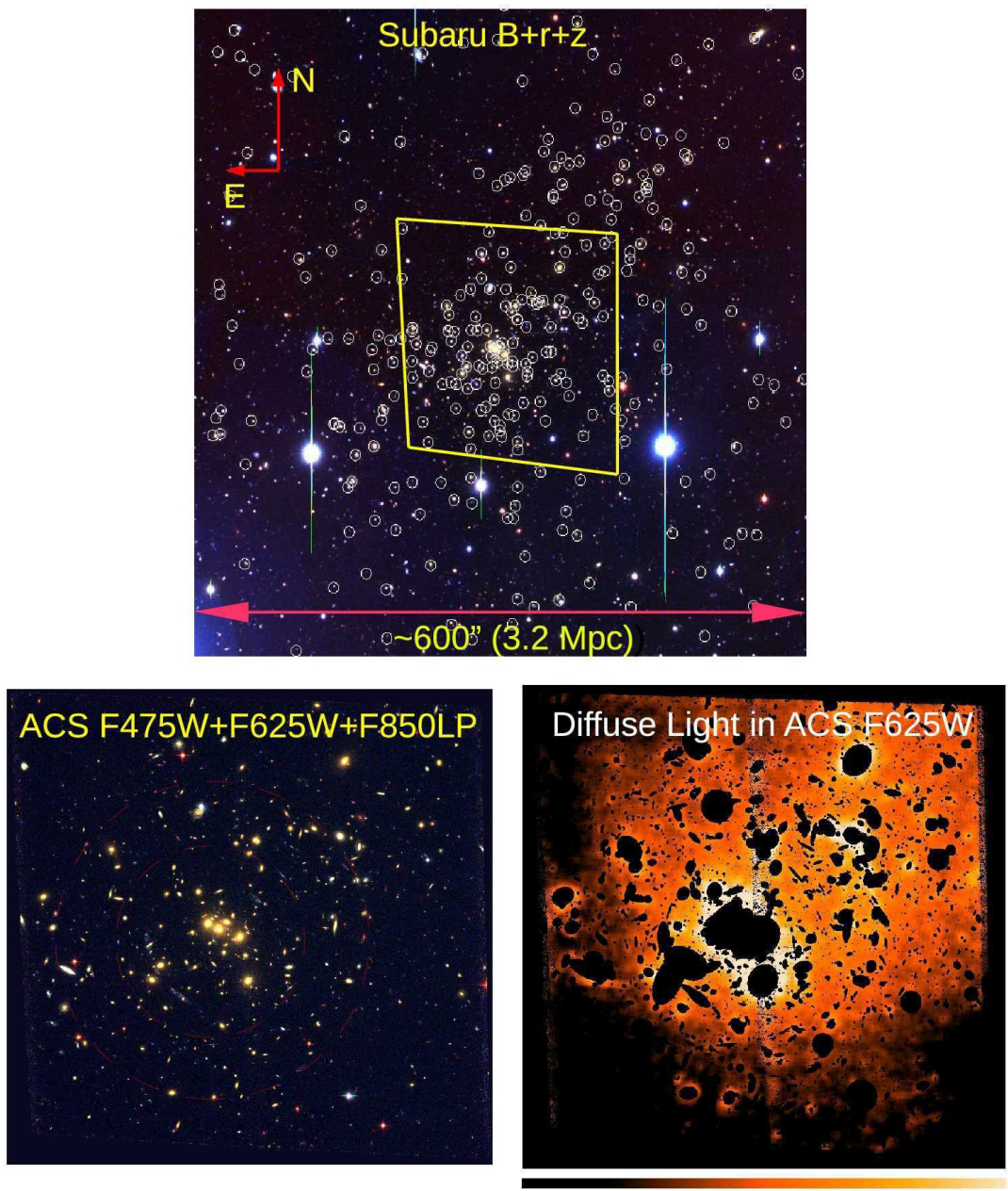}
\caption{ACS and Subaru color images of Cl~0024+17 and two-dimensional ICL map. The top panel shows
the central $10\arcmin\times10\arcmin$ region of the color-composite Subaru image. 
White circles denote spectroscopic cluster members ($0.37<z<0.42$).
The yellow
outline illustrates the $3\arcmin\times3\arcmin$ region observed with ACS. The ACS color-composite image is shown separately
in the lower left panel. The two red dashed-circles delineate the $r\sim40\arcsec$ and $r\sim70\arcsec$ 
radii, across which a sudden change of slope in the ICL profile
occurs. The lower right panel displays the diffuse light in ACS F625W image. This two-dimensional
ICL image is created by masking out astronomical objects, subtracting bright stars, and then 
applying $3\arcsec\times3\arcsec$ box median-smoothing. The color scale is linear. The diffuse
light extends towards the northwestern substructure indicated by the cluster members and the mass reconstruction of Kneib et al. (2003).
\label{fig_ICL_2d}
}
\end{figure}

\begin{figure}
\includegraphics[width=8cm]{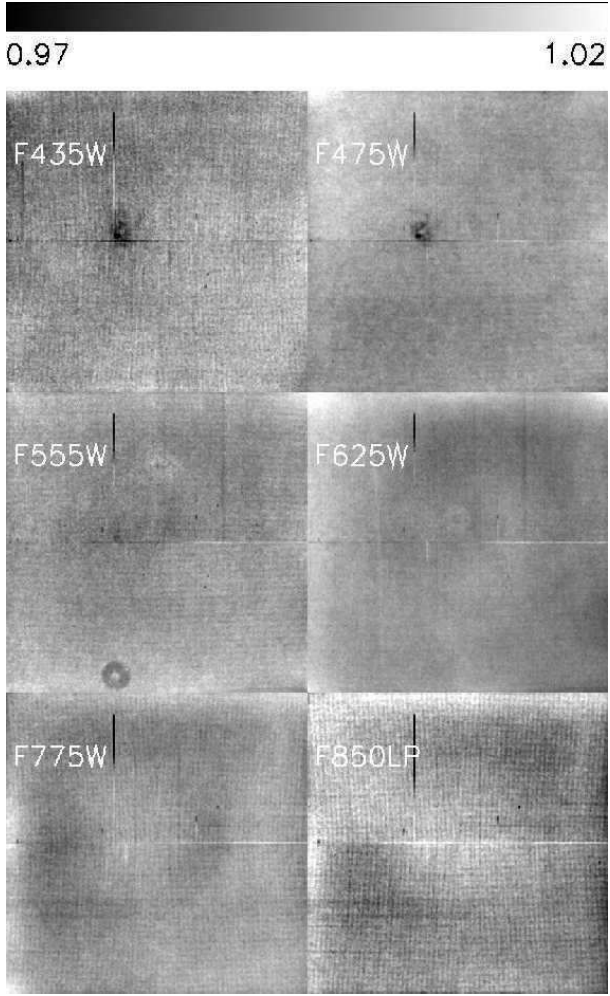}
\caption{Residual sky flats in the six ACS/WFC broadband filters.
These flats are
created from the pipeline-reduced FLT files, and should reveal any residual features that
are not corrected by the pipeline LP-flats (see text for details in the procedure).
The displayed gray scale is linear.
The rms residual is $\sim0.5$\%. The most outstanding
large scale feature is the donut-like pattern particularly clear in F775W and F850LP.
This residual feature may arise from the limited accuracy of the 
polynomial interpolation in the L-flat creation or from the difference of the spectra
between the sky and the 47 Tuc stars as suggested by Pavlovsky et al. (2005).
One of the most conspicuous small scale features is the gridlike pattern. We verified that
this is not an artifact of the 32$\times$32 box median smoothing that we apply before stacking
is carried out. Although we suspect that the feature might be associated with the CCD fabrication procedure,
the exact cause of this $\sim60$ pixel-scale feature is still under investigation.
In addition, the residual flats reveal some new dust-moats and under-sensitive CCD columns.
\label{fig_flat_all}
}
\end{figure}

\begin{figure}
\plotone{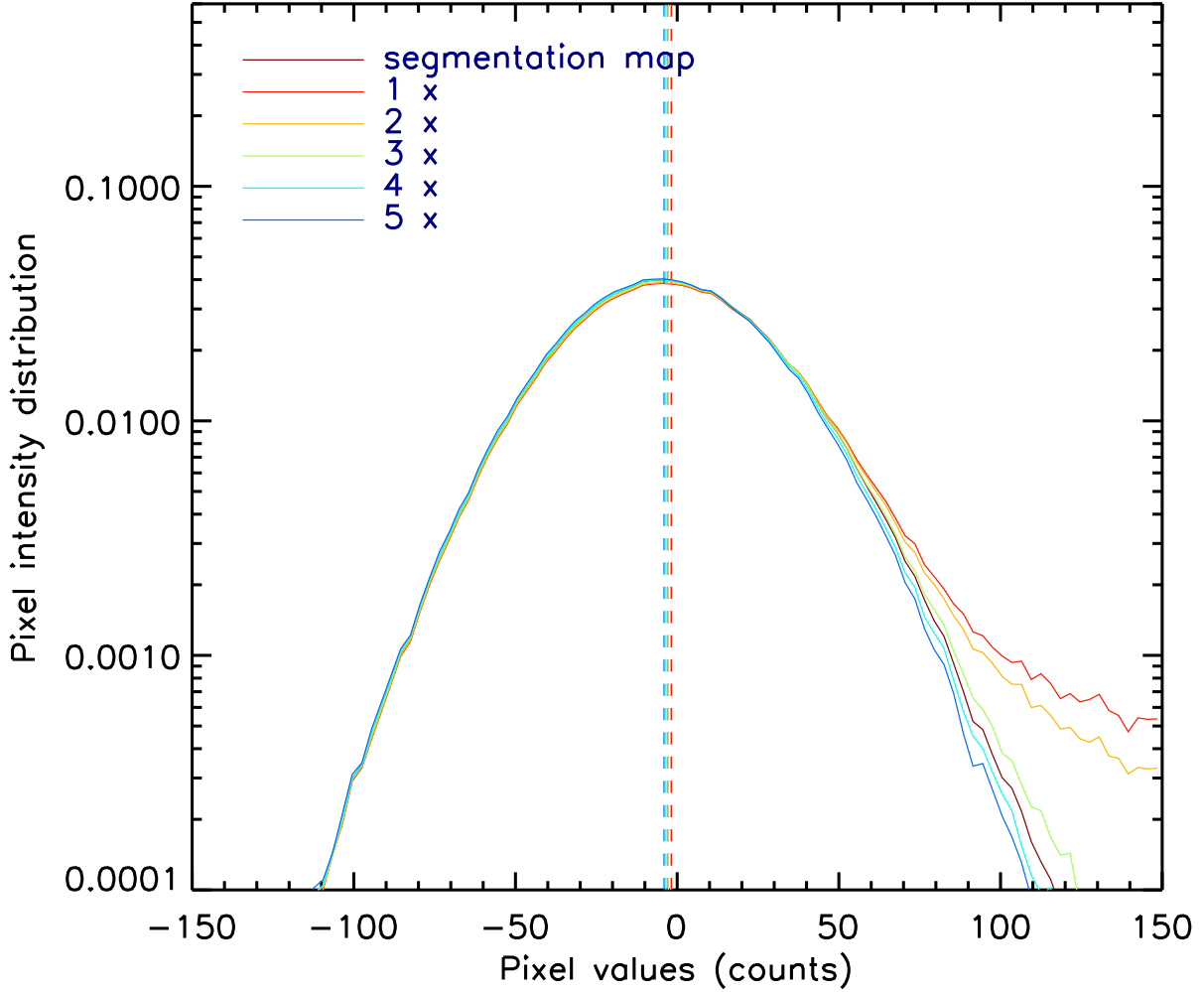}
\caption{Experiment with different masking sizes. We investigated the effect of the masking
size on the centroids of the Gaussian curves by varying masking sizes and repeating the measurements.
As an example, shown here is the result from the $r\sim80\arcsec$ bin in F775W.
The different Gaussian curves represent the results from different masking schemes. The vertical
dashed lines show the corresponding centroids.
We observe that the centroid converges when
the major and minor axes of the masking ellipse are three times or greater than SExtractor measurements.
We conservatively selected five times the values given by SExtractor in defining our masking ellipses throughout
the analysis.
\label{fig_masking}
}
\end{figure}

\begin{figure}
\plottwo{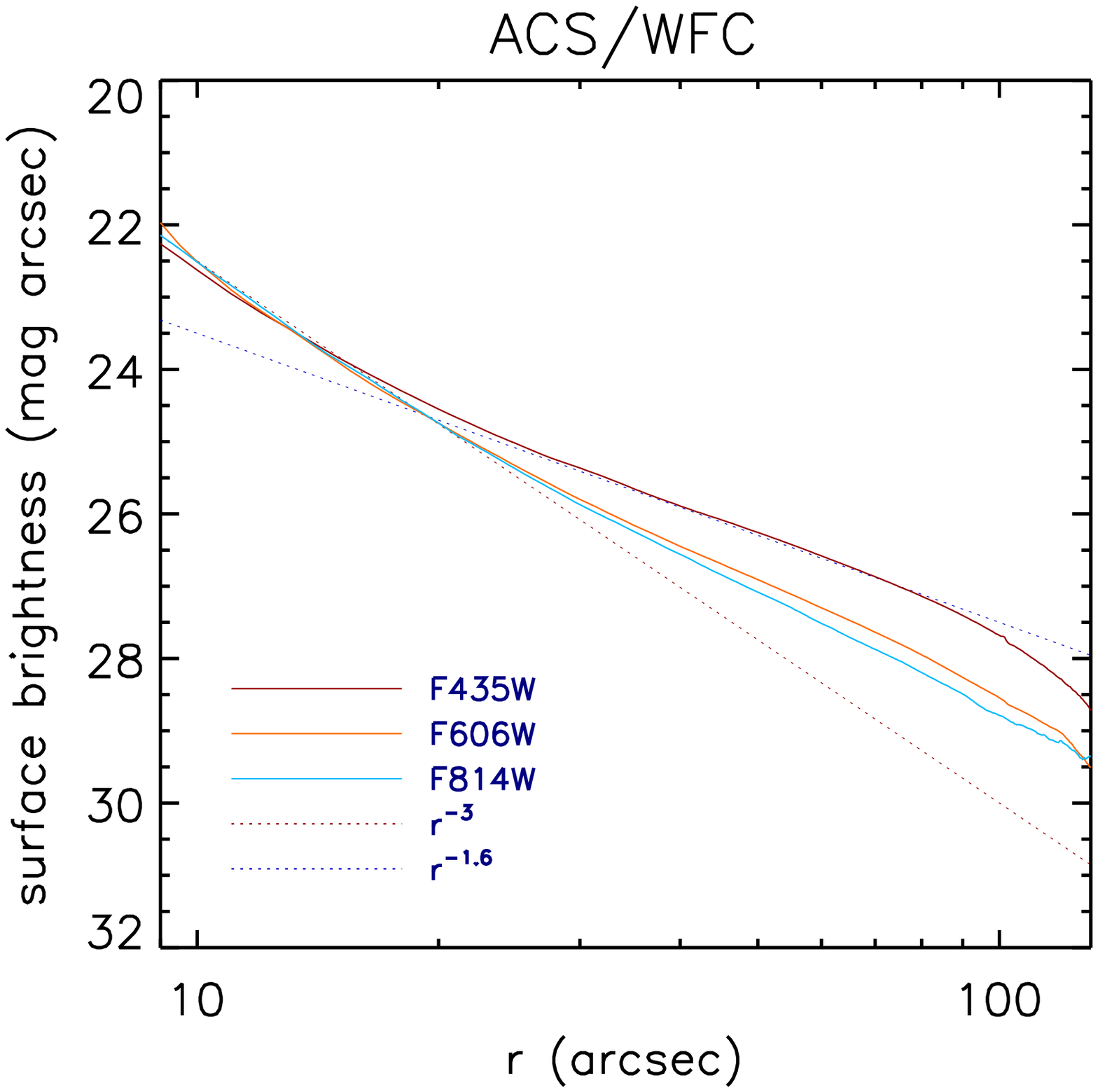}{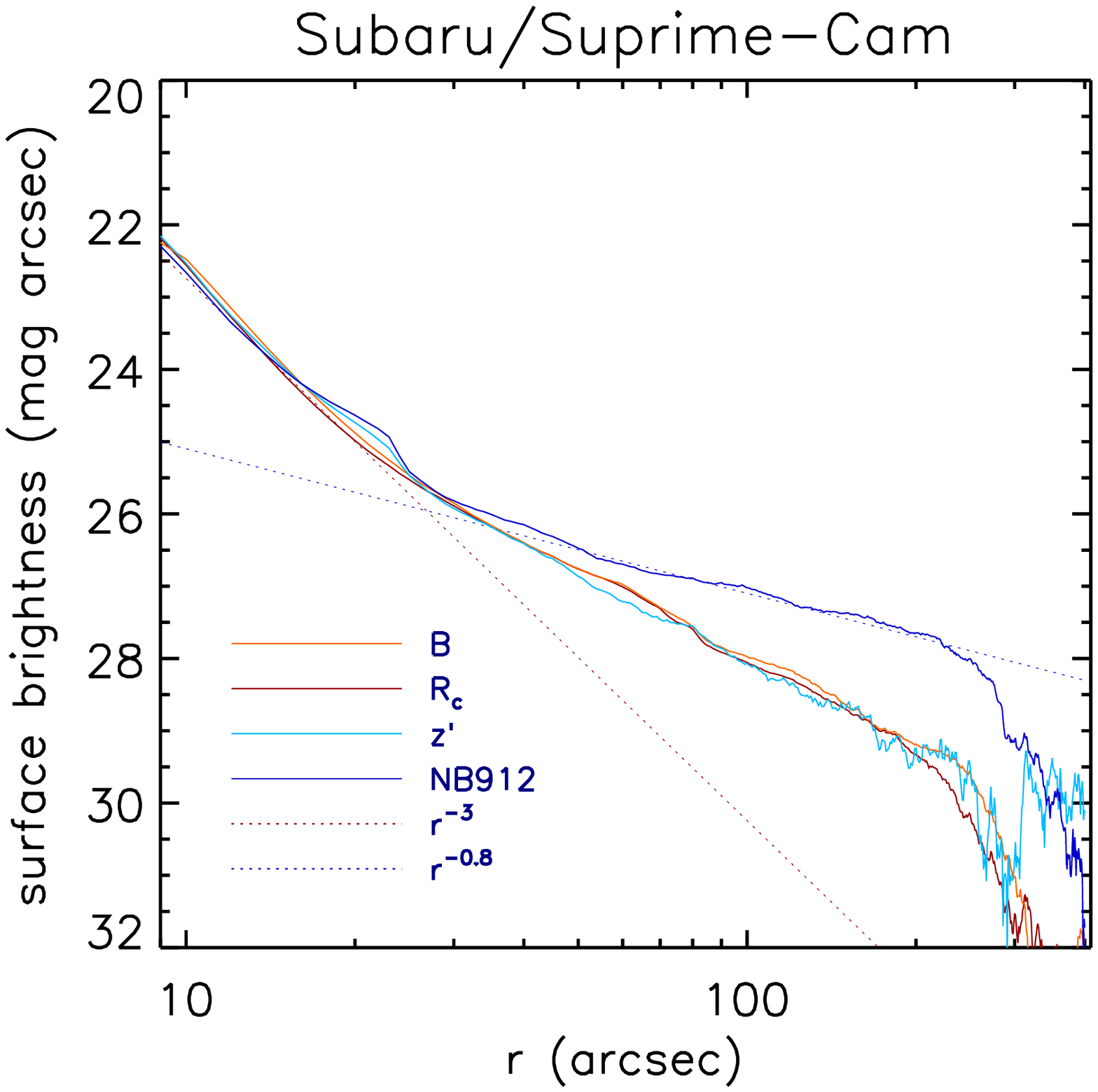}
\caption{Extended PSF wings of ACS/WFC (left) and Subaru/Suprime-cam (right).
The normalization is chosen in an arbitrary way so that the profiles from
different filters approximately overlap at $r<20\arcsec$. The ACS results are
obtained from the archival observation of the star HD39060 (Beta Pictoris).
The Subaru PSFs were measured from the brightest star ($r\sim8.6$ mag) in the Cl~0024+17 image 
($\sim7\arcmin$) away from the cluster center.
\label{fig_psf_wing}
}
\end{figure}

Within the ACS field, we have only 8 moderately bright stars in the magnitude range of $m_R=16-19$.
We masked out the $r=7.5\arcsec$ circular area, which covers the $\mu_R \sim 27-30~\mbox{mag}~\mbox{arcsec}^{-2}$ 
region for these stars. The region outside this circle is 
PSF-subtracted, and we estimate that
the residual error is $\mu_R\gtrsim31~\mbox{mag}~\mbox{arcsec}^{-2}$.
As can be seen in the top panel of Figure~\ref{fig_ICL_2d}, outside the ACS field (within $\sim2 \arcmin$ from the field boundary) 
there are four very bright ($m_R=11-13$) stars. Among these, the PSF wing of the one in the west 
$(\alpha,\delta)=(00:26:24.25,17:08:18.0)$ is non-negligibly
affecting the southwest corner of the ACS image. Based on the magnitude of the star ($m_r\sim11$), we
expect the surface brightness of the PSF wing at the western edge to be $\mu_R\sim28~\mbox{mag}~\mbox{arcsec}^{-2}$.
The residual error after PSF subtraction is estimated to be $\mu_R\gtrsim31~\mbox{mag}~\mbox{arcsec}^{-2}$, and
thus should not affect our ICL measurement.

The lower right panel of Figure~\ref{fig_ICL_2d} displays the diffuse light of the cluster in the ACS F625W image.
The galaxies and the stars are masked out/subtracted as described above. The image shown here is obtained
after median-smoothed with a box size of $\sim3\arcsec \times 3\arcsec$. The color bar represents the intensity
on a linear scale. The ICL distribution of CL0024+17 is somewhat asymmetric and extends northwest toward the secondary galaxy 
number density peak, which lies just outside the ACS field.

For Subaru images, our PSF model extends to $r\sim 5\arcmin$, and we subtracted all $m_R\lesssim17$ mag stars while 
masking out the central region with magnitude-dependent radius. In the region overlapping the ACS field, 
a care was taken in choosing masking radius so that the estimated residual surface brightness error outside the masking aperture
is below the $\sim30~\mbox{mag}~\mbox{arcsec}^{-2}$ level.

\subsection{Background Level Determination \label{section_background}}

The accuracy in the estimation of background sky level affects the
fidelity of the ICL profile at large radii, and is one of the most significant sources of systematic errors
in the current analysis.

For Subaru data, after we removed residual sky gradients and bright stellar profiles 
(\textsection\ref{section_subaru_reduction} and \textsection\ref{section_masking}), we determined
the background level from the $4\arcmin<r<10\arcmin$ annulus.
The inner radius of the annulus was determined by creating the radial profile
of the surface brightness and then locating the region where the profile starts to flatten.
The uncertainty of these background measurements should be dominated by large scale flatfielding errors
and residual sky gradients because the statical noise is only at the $\sim10^{-3}$ \% level of the sky.
We estimated the errors in the background level measurement by subsampling the sky within the $4\arcmin<r<10\arcmin$ annulus.
Table 5 displays these values in terms of fraction of the sky and the corresponding surface brightness level.

The small field of view of the ACS image does not allow us to obtain the background level far from the cluster center.
If the ICL level within the ACS field is still non-negligible, the background level directly
measured within the ACS images biases the ICL level artificially low.
We considered performing photometric transformation from Subaru to ACS and estimating the corresponding
surface brightness level of the ICL (thus indirectly determines the background levels) in ACS. 
However, it is difficult to prove that the two instruments in very different environments allow us
to match the background level in this way.
Therefore, we chose to estimate the background level still within the ACS field while minimizing the
ICL contamination in the following way.

Inspection of the two-dimensional images of the Cl~0024+17 diffuse light 
reveals that the two-dimensional ICL distribution is highly
asymmetric around the center. In case of F625W (e.g., lower right panel of Figure~\ref{fig_ICL_2d}) 
the surface brightness in the northwestern corner is 
$\mu\simeq27~\mbox{mag}~\mbox{arcsec}^{-2}$ whereas it is $\mu\gtrsim31~\mbox{mag}~\mbox{arcsec}^{-2}$ in both the
southeastern and the southwestern corners. Because this high S/N pattern is also supported by other
images of both ACS and Subaru, the two-dimensional feature is believed to represent
the intrinsic distribution of the ICL. In addition, the spatial distribution of the cluster galaxies 
is similar to this two-dimensional ICL map (top panel of Figure~\ref{fig_ICL_2d}).
Therefore, we argue that if we measure the background level at $r>80\arcsec$ from the ACS image while excluding the northwest region, we
can minimize the influence of the ICL.
This argument is further supported by our
experiment with the Subaru image, which shows that the background level measured in this way (i.e., cropping the $3\arcmin\times3\arcmin$ region
to simulate the ACS field) is in good agreement with the result measured from the $4\arcmin<r<10\arcmin$ annulus 
within the $\mu\simeq30~\mbox{mag}~\mbox{arcsec}^{-2}$ level.

\subsection{X-ray Data Reduction}

For the $Chandra$ X-ray analysis, we followed the procedures described in J07, using the $Chandra$ Interactive Analysis of
Observations (CIAO) software version 3.3 and the Calibration Database (CALDB) version 3.2. 
The XMM-$Newton$ data
(taken on January 2001 for a total integration of 52.1 ks, 52.1 ks, and 48.3 ks for MOS1, MOS2, and PN)
were retrieved from the XMM-$Newton$ Science Operations Centre\footnote{The data are available at http:\/\/xmm.esac.esa.int} and processed with the Science Analysis Software (SAS)
version 7.1.2. We combined the three instrument data and applied the exposure map to obtain an exposure-corrected image.
We detected point sources in the $Chandra$ and used the results to mask both the $Chandra$ and XMM-$Newton$ exposure-corrected images
before we measured the final X-ray surface brightness profile.

\section{ICL ANALYSIS \label{section_icl_analysis} }

\subsection{Measurement of the Radial Profile \label{section_measurement}}

We adopt the statistical approach of Uson et al. (1991) in measuring the radial
ICL profile of the cluster. In
their analysis of Abell 2029, they constructed histograms from the pixels in 
radial bins and demonstrated that for each bin the position
of the Gaussian peak in the pixel intensity histogram is a fair indicator of the ICL levels.
The skewness of the Gaussian, mainly
caused by diffuse light from wings of astronomical objects, is a potential source of bias
in this measurement. Uson et al. (1991) report that the bias introduced by this contamination
is about 0.2\% of the sky level at one core radius of the cluster if objects are left unmasked.
As discussed in \textsection\ref{section_masking}, our ``5 $\times$" masking scheme reduces
the effect below the 0.02\% level of the sky.

\begin{figure}
\plotone{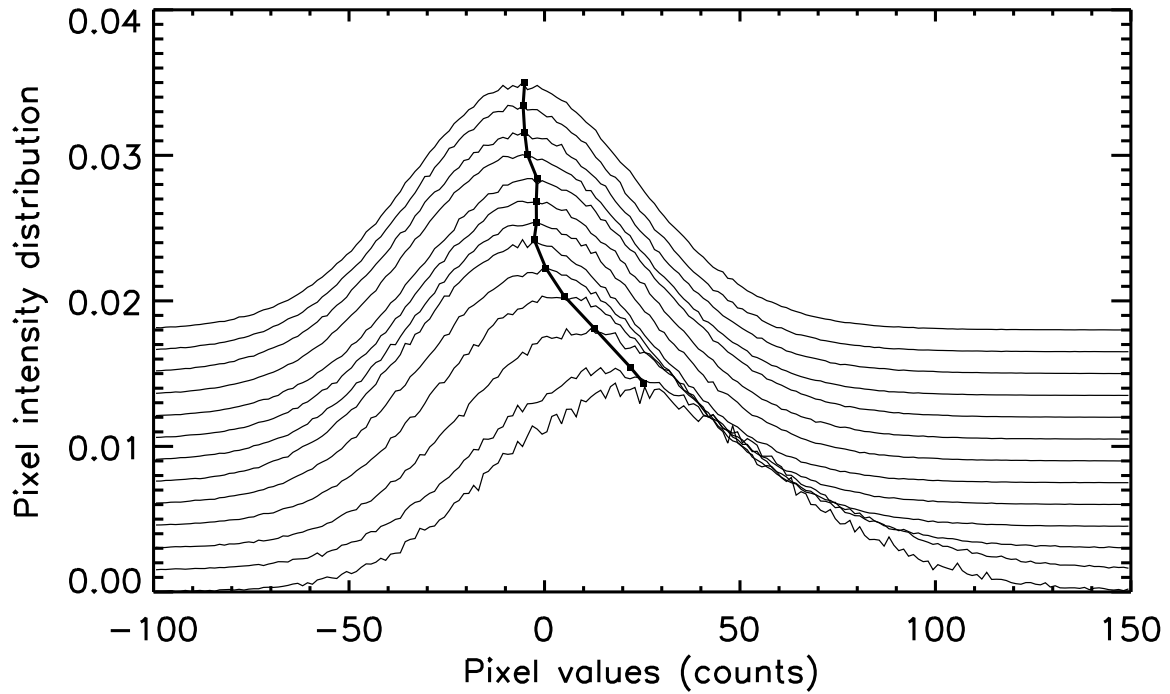}
\caption{Pixel intensity distribution in the ACS/WFC F775W image. Objects were masked and
bright stars were subtracted before the statistics was evaluated.
Each curve represents the pixel intensity distribution in each radial bin.
The bottom curve represents the data in the $r=8\arcsec$ bin. To improve readability, 
we displace the histograms vertically. The top
curve corresponds to the data in the $r=104\arcsec$ bin. We adopt the center of the Gaussian peak
as the ICL level of the bin (filled square and thick solid line). The uncertainties
in determining the centroid of the Gaussian curves after including photon noise and flatfielding errors 
are $\sim0.5$ counts, less than the
size of the filled square symbols.
\label{fig_many_histos}
}
\end{figure}

To illustrate our measurement scheme, we show in Figure~\ref{fig_many_histos} the histograms that we
obtain in F775W after masking objects and subtracting stars. We set up radial bins out to $r\sim100\arcsec$ 
($\sim530$ kpc) at an interval of $\Delta r=8\arcsec$.
The bottom curve represents the data in 
the $r=8\arcsec$ bin. We displace
the subsequent histograms vertically to improve readability. The top curve corresponds
to the data in the $r=104\arcsec$ bin. The filled squares and thick solid line
trace the location of the Gaussian peaks. The uncertainties in determining the centroid
of the Gaussian curve after including photon noise and flatfielding errors are $\sim0.5$ counts 
or $\mu\simeq30~\mbox{mag}~\mbox{arcsec}^{-2}$, smaller than the size of the filled square symbol.
We choose the geometric center of the dark matter ring
as the origin of the radial bin. The geometric center of the dark matter ring is $\sim10$\arcsec offset toward south
from the peak of the X-ray emission (also the location of one of the brightest cluster galaxies).

\begin{figure}
\plottwo{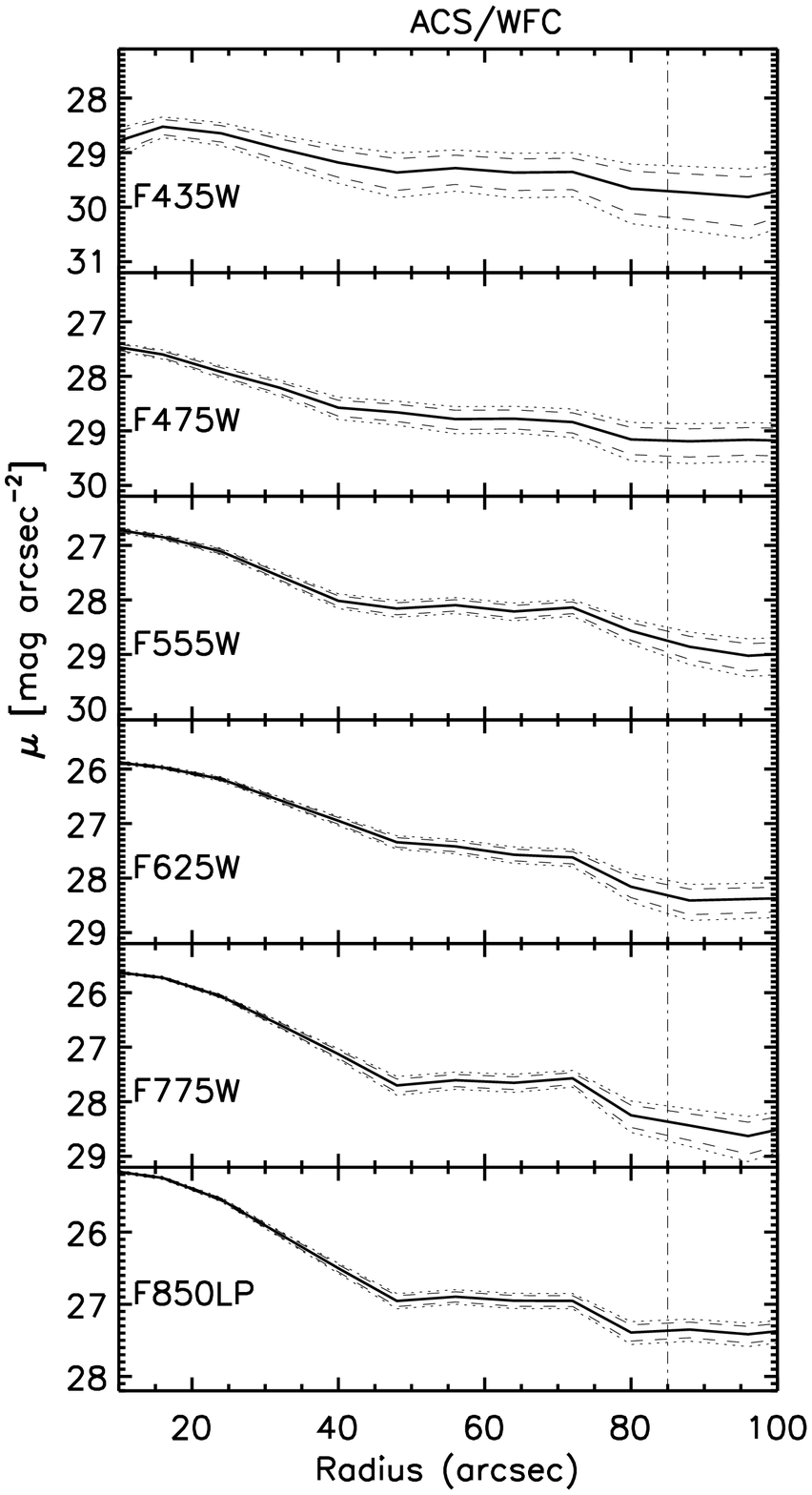}{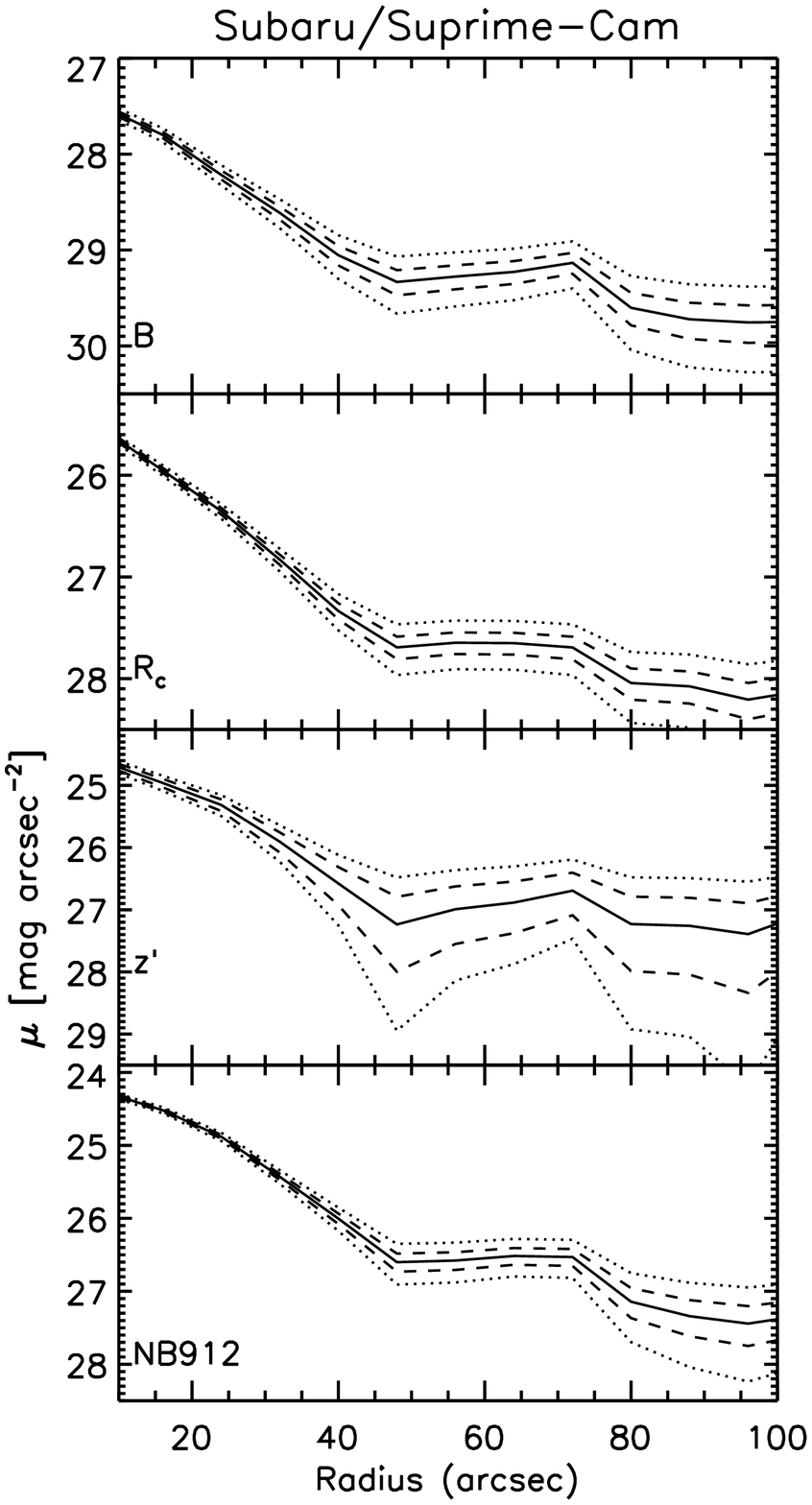}
\caption{ICL profile in Cl~0024+17. The thick solid lines represent the surface brightness while the
dashed (dotted) lines show the 1-$\sigma$ limits without (with) errors in background level
determination included.
In case of ACS, we arbitrarily assume that the error in background level measurement is at the 0.1\% of the
sky level.
Because of the limited field of view, ICL measurement from the ACS data is obtained from an incomplete circle at $r\gtrsim85\arcsec$
(dot-dashed). 
\label{fig_icl}
}
\end{figure}

Figure~\ref{fig_icl} shows ten ICL profiles that we measure from both ACS and Subaru data for all available passband data.
The thick solid lines represent the surface brightness while the
dashed (dotted) lines show the 1-$\sigma$ limits without (with) including background level measurement errors.
In case of ACS, we arbitrarily assume that the error in background level measurement is at the 0.1\% of the
sky level (the statistical error is negligible).
The Subaru profiles have much larger errors mainly because the sky levels are higher. Especially, in the $z^{\prime}$
image the sky level is nearly 3 magnitudes brighter than in the ACS F850LP image, and thus the 0.05\% flatfielding
error is translated into $\mu\sim27.5~\mbox{mag}~\mbox{arcsec}^{-2}$ whereas the 0.1\% flatfielding error
in the ACS F850LP corresponds to $\mu\sim30~\mbox{mag}~\mbox{arcsec}^{-2}$.

Despite the filter-by-filter variation in significance, it is remarkable that all profiles in Figure~\ref{fig_icl} share
some conspicuous common features. 
The surface brightness in every profile decreases rapidly from the cluster center to $r=40\arcsec-50\arcsec$. Then,
it maintains its level out to $r\sim70\arcsec$, where it starts to decrease again for increasing radius. The changes
of the slopes at the two locations are nearly discontinuous, somewhat more sudden than the corresponding changes in mass
slopes of J07; the significance of the feature should be evaluated using the inner error lines (dashed), which
exclude the background level measurement errors.
This striking consistency across different filters and instruments is
the evidence against the possibility that the features come from any residual systematic errors.

The comparison of the ACS and Subaru results at the large radii help us to assess the degree of the ICL contamination in the background
level for ACS. Although the difference in throughput curve between the instruments complicates the detailed comparison,
there is no apparent indication that any significant systematic underestimation of the ICL level (or overestimation of the background level) is introduced
in the ACS results; the ICL levels in corresponding filters between the two instruments are consistent within the errors.

\subsection{Comparison of ICL Profile to Mass Profile \label{section_comparison}}

The peculiar mass density profile of Cl 0024+17 in J07 is characterized by
the steep decline out to $r\sim50\arcsec$ ($\sim265$ kpc), the slow rise from $r\sim50\arcsec$ ($\sim265$ kpc) to $\sim75\arcsec$ ($\sim397$ kpc),
and the modest decrease at $r\gtrsim75\arcsec$. The turnaround at $r\sim75\arcsec$ ($\sim397$ kpc) appears as
the dark matter ring in the two-dimensional mass reconstruction. 
The ICL profiles in Figure~\ref{fig_icl} clearly signals these features of the J07 mass profile.
Although the significance varies across filters and instruments, it is evident that all profiles
change their slopes conspicuously at both $r\sim50\arcsec$ and $r\sim72\arcsec$ in the similar
way as in the mass profile.
Considering the size of the uncertainty (we exclude the background level uncertainty here because it does not
affect the significance of the relative variation), the change of the profile slopes
is significant in every panel except for the ACS F435W data, for which the major source uncertainy comes from a low ICL-to-sky ratio.
Of course, the most remarkable
aspect of the results is the observation that all profiles closely resemble one another, which strongly rules out
the possibility that the source of these features originate from uncorrected instrumental systematics.

\begin{figure}
\plotone{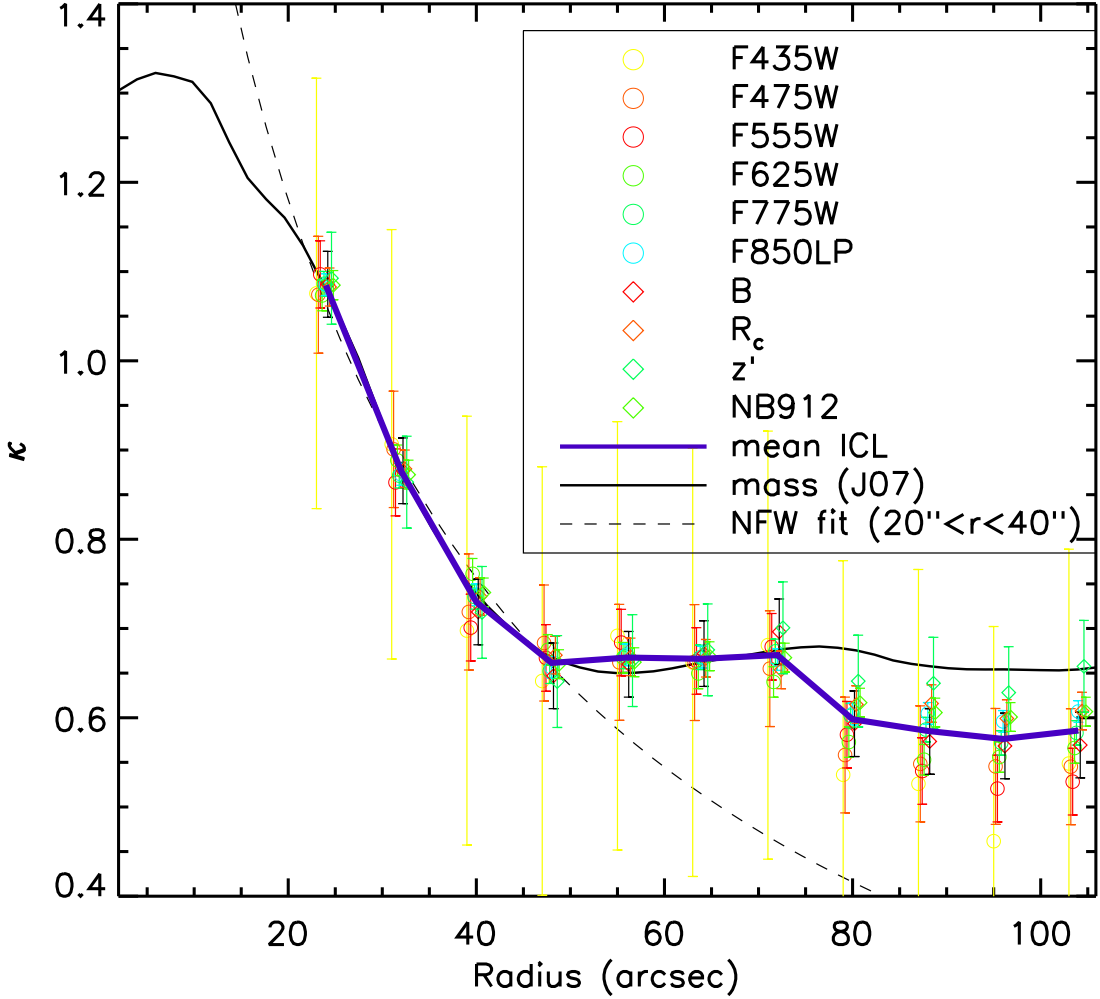}
\caption{Comparison of the ICL profiles with the mass density profile. The ICL profiles from nearly all the filters
reveal the critical features of the mass profile: the steep decline at $r\lesssim50\arcsec$, the slow rise at $50\arcsec \lesssim r \lesssim 75\arcsec$, and
the turnaround at $r\sim75\arcsec$. However, we note that the ICL profiles decrease faster after their turnover at
$r\simeq72\arcsec$ than the mass profile of J07. We display the NFW fit result (obtained by forcing the profile to match
the mass at $20\arcsec<r<40\arcsec$) in order to illustrate how much both the cluster mass and ICL profiles deviate
from this analytic model frequently used to describe relaxed clusters.
\label{fig_masstoicl}
}
\end{figure}

In Figure~\ref{fig_masstoicl} we overplot these ICL profiles on top
of the mass profile of J07 after rescaling of the flux via the following
form: $I^{\prime}(r) =a \times I(r) +b$. If intracluster stars are truly
poissonian tracers of underlying dark matter, a strict comparison would
require no intercept in the transformation. However, considering
the uncertainty of the background level determination in ICL measurement,
the uncertainty of the mass profile caused by mass-sheet degeneracy, and
the physical mechanism that ICL might be more centrally concentrated, we allow
the intercept to vary in the fitting procedure. Furthermore, we
limit the fitting range to the $20\arcsec<r<75\arcsec$ region because
at $r<20\arcsec$ the ICL measurement is hampered by severe masking
of bright cluster galaxies, and at $r>75\arcsec$ the 
ACS data become progressively incomplete toward large radii. Also, for 
some filters ICL profile measurements are sensitive to the background 
estimation in this regime.
The black solid line is the mass profile of J07. The open circles
denote the results from the different filters in ACS whereas the open diamonds
represent the case for the Subaru images. The thick purple line delineates
the mean ICL profile. 

This direct comparison more graphically supports the above claim that
the ICL of Cl~0024+17 mimics the peculiar mass profile. 
The mean ICL profile shows a sudden change of slope at $r\sim50\arcsec$ and
$r\sim72\arcsec$ as does the mass profile. In addition, the difference
among the results from different filters is remarkably small. Because the
agreement is excellent over a large range of radii, this consistency
should not be attributed to the rescaling of the profile.

We note however that the agreement between the ICL and mass profiles
degrades at $r\gtrsim72\arcsec$. Apparently, the ICL profile changes
its slope more suddenly at $r\sim72\arcsec$ than the mass. The difference
among the ICL profiles is also slightly larger in this regime. 
It is tempting to view this as indicating that the ICL distribution is
more centrally peaked than the cluster mass distribution.
However,
one should use a caution in interpreting the difference.
The ACS data do not provide complete azimuthal statistics at 
$r\sim80\arcsec$. Also, the ICL level in this range is sensitive
to the background level determination. 

The dashed line is an NFW fit to the mass profile
using the value in the $20\arcsec<r<50\arcsec$ range. 
Both the cluster mass and ICL profiles deviate from this analytic 
description, which is frequently used to model halos of
relaxed clusters.

\subsection{Is the Peculiar ICL Structure Local?}

The consistent detection of the peculiar ICL feature of Cl~0024+17 from both ACS and Subaru 
data provides
evidence that the feature does not come from any residual calibration errors. 
And the similarity between the ICL and mass profiles adds further to the credibility of the detection.
However, one can raise a concern that our masking might miss a few low surface brightness
objects that fortuitously at a similar distance from the cluster center. While
we consider that this is very unlikely given the significance and the scale of the features,
it is still useful to devise a test demonstrating that the features indeed come from a two-dimensional, quasi-symmetric diffuse light distribution, not from any localized
concentration.

The most straightforward proof would be the revelation of the discussed feature 
in the two-dimensional ICL map.
However, given the low contrast of the structure, it is extremely difficult, if not 
impossible, to produce
a convincing diffuse light map of the cluster. The pixel histograms in 
Figure~\ref{fig_many_histos} that we
discussed in \textsection\ref{section_measurement} help us to quantitatively 
understand why it is
next to impossible to see the feature in a two-dimensional map. As discussed already, 
the location of the Gaussian
curve centroid is the indicator of the ICL level. The measurement leads to an uncertainty small enough 
($\sim30~\mbox{mag}~\mbox{arcsec}^{-2}$)
for us to give high significance to the relative variation of the centroids as a function 
of radius. However, the width of the Gaussian curve itself (i.e., 
the intrinsic pixel intensity distribution) is much larger than the relative variation 
of the centroids (i.e., $\sim70$ counts versus $\sim2$ counts), which makes it futile
to attempt to detect the feature in a two-dimensional map.

\begin{figure}
\includegraphics[width=8cm]{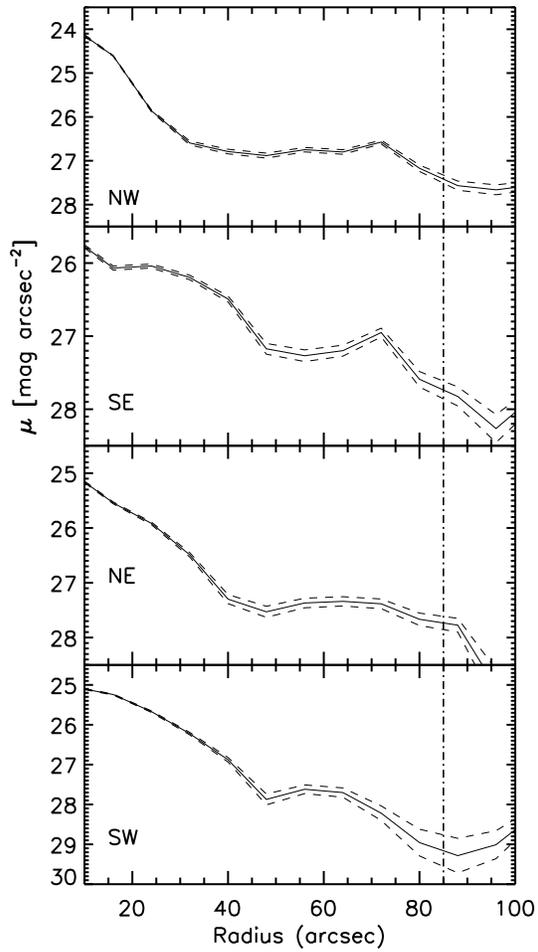}
\caption{Same as in Figure~\ref{fig_icl} except here we measure the ICL profile in each quadrant of the ACS F775W image.
Despite the quadrant-to-quadrant variations, the signal of the peculiar ICL structure is present in all four quadrants.
The dashed-line does not include the uncertainty from background measurement.
\label{fig_icl_f775w}
}
\end{figure}

A visual inspection of the two-dimensional ICL map (the lower right panel of 
Figure~\ref{fig_ICL_2d}), nevertheless, 
reveals that there is some localized diffuse light in the northwestern quadrant; 
it appears that the diffuse light
in this region is associated with several galaxies approximately on the $r\sim72\arcsec$ circle. 
Hence, it is
important to examine how significantly the ICL profile that we presented in 
\textsection\ref{section_measurement} is
affected by this apparent substructure. To address this issue, we divided the cluster field into four azimuthal
regions and measured an ICL profile in each quadrant. The result from the F775W image 
is presented in Figure~\ref{fig_icl_f775w}.
Although there exists a quadrant-to-quadrant variation, the features seen in Figure~\ref{fig_icl}
is also observed in each quadrant. Most of all, the ICL level stops
decreasing at $r=40\arcsec\sim50\arcsec$ and increases out to $r\sim70\arcsec$ in each panel.
The sudden change of slope at $r\sim72\arcsec$
is the strongest in the NW and SE quadrants. The profile from the NE quadrant shows a weak change of
the slope at this location whereas the slow rise of the profile between $40\lesssim r\lesssim 70\arcsec$ is still clear.
The SW ICL profile possesses a sudden change of slope at $r\sim50\arcsec$ and the surface brightness is increasing
out to $r\sim65\arcsec$. For this quadrant, the change of slope at $r\sim72\arcsec$ is subtle, which suggests
that the ringlike structure might not be strong there. Nonetheless, the interpretation 
is difficult because the surface brightness level at this location is very low ($\mu\gtrsim29~\mbox{mag}~\mbox{arcsec}^{-2}$)
and thus the error is dominated by systematics\footnote{The total amount of the ICL flux becomes
negative when we subtract the background level. Therefore, we arbitrarily adjusted the background value to prevent it}.

Therefore, from the above experiment we conclude that the ICL profile structures resembling the peculiar mass profile in J07
do not come from any localized substructure, but from a two-dimensional, quasi-symmetric 
distribution of diffuse light in Cl~0024+17.

\subsection{ICL Color versus Galaxy Color \label{section_icl_color}}

The color of ICL relative to the color of cluster galaxies potentially
constrains the epoch at which most of the intracluster
stars were stripped from the cluster galaxies. 
The literature roughly converges on the observation that ICL colors are approximately 
consistent with the cluster galaxy colors, which suggests that intracluster
stars may be just as old as cluster galaxies. 

\begin{figure}
\plottwo{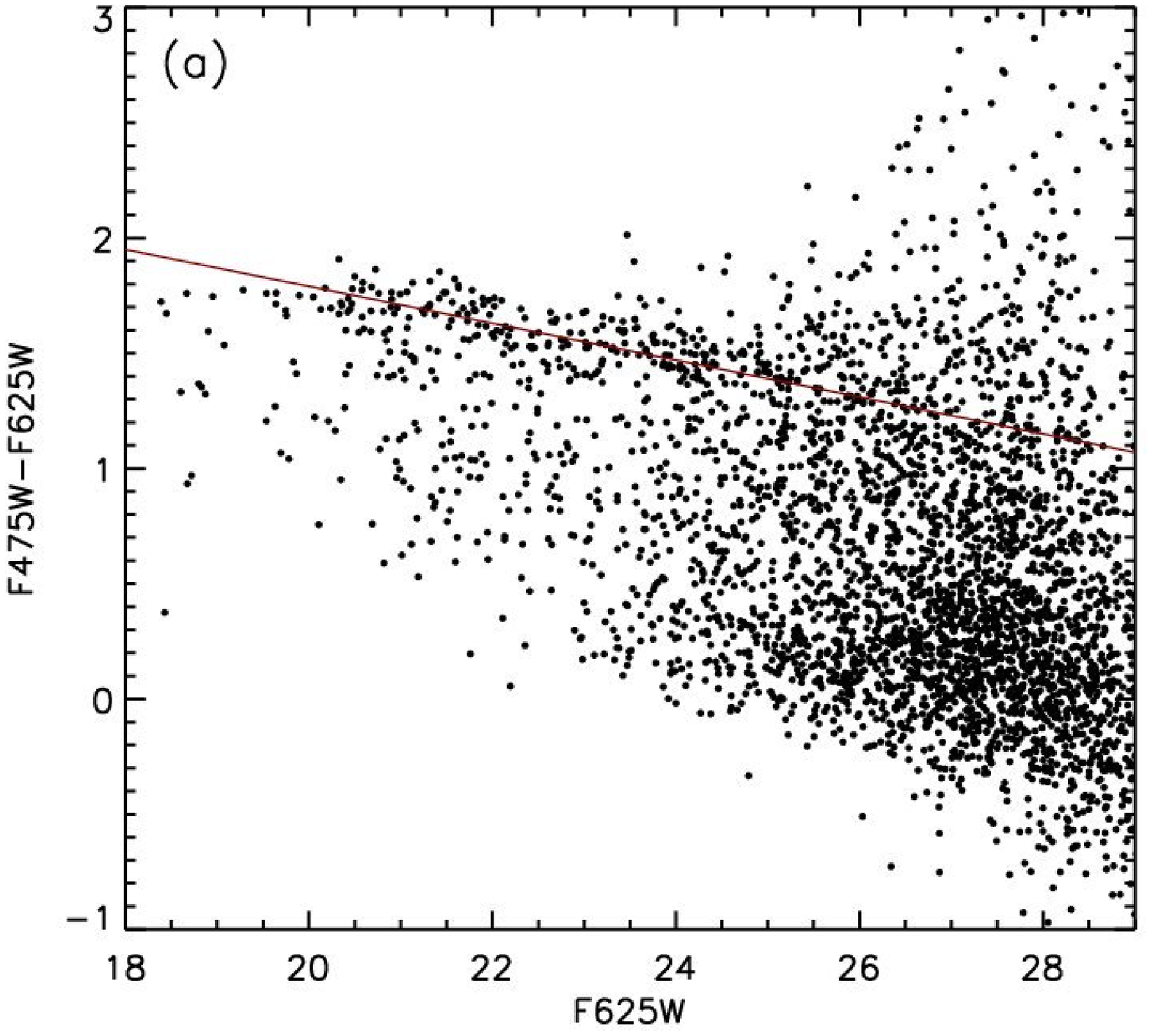}{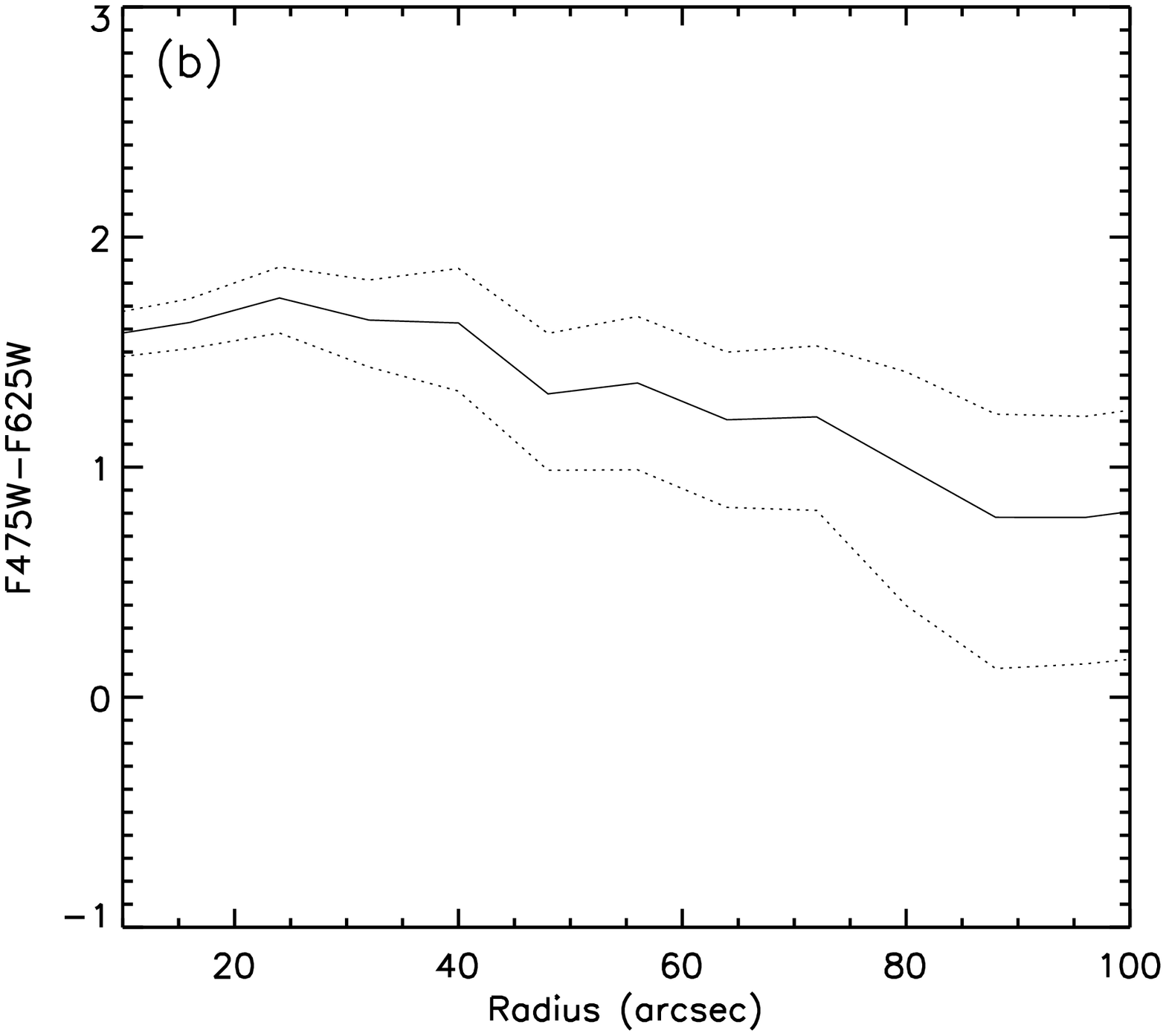}
\caption{Galaxy color versus ICL color. (a) Color magnitude diagram in the ACS
field. We used SExtractor's isophotal magnitude to compute the galaxy colors
while the F625W magnitude in the X-axis is SExtractor's MAG\_AUTO. The red solid
line approximately indicates the location of the cluster red-sequence.
(b) ICL color profile as a function of radius. The dotted lines represent the
1-$\sigma$ uncertainty including errors in photon statistics, flatfielding
inaccuracy, and background level determination. We assume that the error in the background level
determination is at the 0.1\% level of the sky.
\label{fig_icl_color}
}
\end{figure}

Considering the location of the 4000~\AA~break at $z=0.4$ in the observed
frame, we find that a F475W-F625W color gives good contrast between
the cluster red sequence and the rest of the population in the field.
The color magnitude diagram in Figure~\ref{fig_icl_color}a with this
filter combination clearly shows the red sequence of the cluster in the ACS
field (the red solid line approximately delineates the location). We 
display 
in Figure~\ref{fig_icl_color}b the F475W-F625W color profile
of the ICL.  At large radii, the accuracy highly depends on how well we
can determine the net amount of ICL. Therefore, the biases  in the 
background level determination cannot be neglected here. We assumed that
this bias is at the 0.1\% of the sky level.

We find that the ICL color at small radii ($r\lesssim40$) is consistent 
with that of the cluster red sequence while
the color profile tends to become bluer
with increasing radius.  This trend is in agreement with the general expectation
that cluster galaxies are bluer at larger cluster-centric radii.
Robust interpretation of Figure~\ref{fig_icl_color}b would be possible if we use
a population synthesis model assuming several different star formation histories.
While we acknowledge the importance of this study in determining the
age of the progenitor population of the ICL, this extensive
analysis is beyond the scope of the current paper.
However, one possible interpretation of Figure~\ref{fig_icl_color} is that
for the Cl~0024+17 cluster the observed ICL might be the result
of on-going stripping of stars (e.g., Moore et al. 1999). If dominant fraction of the ICL stars were
stripped a few Gyrs earlier than the current epoch and have evolved passively since then, we would expect
to observe that the intracluster stars are redder than the cluster galaxies
because the cluster galaxies continue to form new stars (e.g., Sommer-Larsen et al. 2005).

\subsection{ICL Fraction \label{section_icl_fraction}}

Together with ICL color, the total amount of light in
intracluster stars relative to the total cluster light is also an important
quantity in helping us to infer the production history of intracluster stars.
In order to identify the total cluster light, we combined the photometric
redshift catalog of J07, and the publicly available spectroscopic catalog
of Moran et al. (2005). 
Figure 3 and 4 of J07 demonstrate that the photometric redshifts
estimated from six ACS filters with Hubble Deep Field North (HDFN) prior 
give highly consistent results with the spectroscopic measurements and
reliably identify the cluster members. 
We replaced the photometric redshifts if their spectroscopic redshifts are known
($\sim140$ objects) and used this merged catalog to select
 the cluster members of Cl 0024+17.  The rest of the procedure is
 identical to the one in our ICL measurement described in
 \textsection\ref{section_measurement} except that 1) we did not
 mask out the cluster members and 2) took the mean value in each
 annulus instead of the location of the Gaussian peak.

\begin{figure}
\includegraphics[width=8cm]{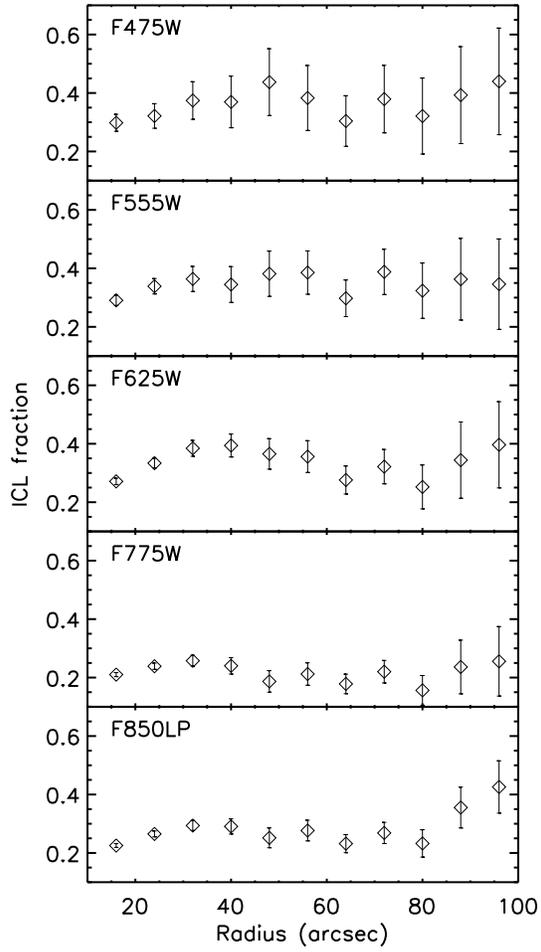}
\caption{Ratio of ICL flux to total cluster flux. We combined the photometric
redshift catalog of J07, and the publicly available spectroscopic catalog
of Moran et al. (2005) to identify the cluster members.
\label{fig_icl_fraction}
}
\end{figure}

Figure~\ref{fig_icl_fraction} shows the ICL fraction measured from the five
ACS filter images; we omit the results for F435W, whose error bars are
substantially larger than the rest.
As in the case for the ICL color measurement,
the ICL fraction at large radii is sensitive to the background level
determination and we again assume 0.1\% of the sky level for the uncertainty.
It appears that the mean level of the ICL fraction somewhat depends on
the used filter. Although this observed trend possesses low significance
at large radii due to the systematics in the background level determination, 
the results at small radii ($r\lesssim40\arcsec$) show that
the ICL fraction is on average higher in blue filters (i..e., F475W, F555W, and F625W). 

In order to estimate the cumulative ICL fraction within an aperture, we need to assume
the behavior of the ICL profile at $r<15\arcsec$, where our masking procedure
left no area for direct ICL measurement. Extrapolating the central value using the trend at $r<40\arcsec$,
we obtain the results in Table 4. The uncertainties are determined from Monte-Carlo simulations 
while the correlations between points are taken into account.
The literature does not converge on the ICL fraction, and perhaps this represents rather
a wide range of values for different clusters than scatters in measurements.
Certainly, the ICL fraction ($\gtrsim30$\%) of Cl~0024+17 that we quote here indicates that
the cluster might be near the high end of the distribution.
Tyson et al. (1998) determined the ICL fraction for Cl~0024+17 to be 15\%$\pm$3\% within
the $rsim100~$kpc region. Because their analysis is based on the Wide Field Planetary Camera 2 (WFPC2)
data, which covers only $\sim50$ \% of the ACS field, we suspect that their background level measured within
the WFPC2 field might be biased high due to the significant ICL level; the ICL profile in our study shows
that the surface brightness level remains high out to $r\sim70\arcsec$.

Given the numerical studies suggesting that most of intracluster
stars become unbound at $z<1$ with no preferred epoch (e.g., Murante et al. 2007),
the high ICL fraction of Cl~0024+17 at $z=0.4$ is somewhat unexpected.
However, if the high-speed encounter scenario of Czoskey et al. (2001)
for Cl~0024+17 is considered, it is plausible that the collision of two equal-mass subclusters 1-2 Gyrs ago 
might have liberated unusual amount of stars in the cluster galaxies. 

\section{INTRACLUSTER MEDIUM PROFILE OF Cl~0024+17}

Although the collisional and dissipative nature makes its spatial distribution often
differ from that of dark matter, intracluster gas distribution in a galaxy cluster 
strongly signals the structure of the cluster potential dominantly
determined by the cluster dark matter. 
Therefore, whether or not one
can find a good spatial agreement between the two results, a detailed comparison often provides
an insight into the dynamical history of the two cluster constituents 
(e.g., Jee et al. 2005; Clowe et al. 2007).

Our previous analysis of the Chandra data of Cl~0024+17 confirmed the unusual X-ray profile of the cluster
originally reported by Ota et al. (2005). The central excess of the X-ray photons only allows
us to describe the profile with a superposition of two isothermal profiles of different $\beta$ parameters. 
Nevertheless, the X-ray surface brightness profile of J07 did not show any features reminiscent of
the reported dark matter profile. 

\begin{figure}
\plotone{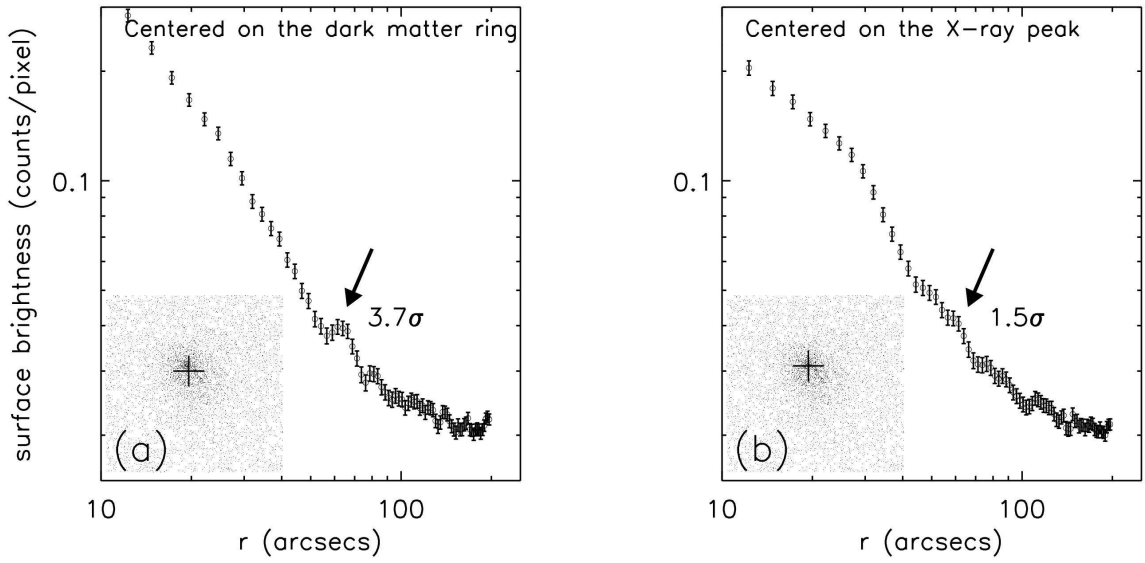}
\caption{X-ray surface brightness profile created from the $Chandra$ data.
(a) When the origin agrees with the center of the dark matter ring, it reveals
an overdensity at $r\sim60"$ (the inset picture shows the point-source removed X-ray image and
the `+' symbol denotes the origin of the radial bin).
(b) When we align the origin to the peak of the X-ray emission (also the location of the mass and luminosity peak), 
the bump in (a) becomes much less prominent.
We let the neighboring bins overlap by $\sim50$\% to suppress the shot noise. 
\label{fig_xray_profile}
}
\end{figure}

However, repeating the J07 analysis with a different choice of the cluster center shows that
there presents a $3.7 \sigma$ bump in the X-ray profile at $r\sim60\arcsec$ as shown 
in Figure~\ref{fig_xray_profile}. The result in the left panel is obtained when
the center of the dark matter ring is chosen as the origin of the radial
bins whereas the the right panel shows
the X-ray profile when the center is placed on the peak of the X-ray
emission as in the J07 analysis. It is worth noting that the distance between these two locations is only
$\sim10\arcsec$. The fact that this small change greatly alters the
significance of the feature indicates that the contrast is low and the
bump is associated not with any unidentified point source, but with
smooth, azimuthal distribution of excess X-ray photons.

The X-ray surface brightness stops decreasing at $r\sim50\arcsec$ and
increases slowly until it suddenly drops again at $r\sim60\arcsec$. The
overall behavior is reminiscent of the features seen in both the ICL and
mass profiles of the cluster. 

Similar to the case in our ICL measurement, this X-ray feature becomes
more credible if a consistent structure is observed in data from a
different instrument.  We present the result from the XMM-$Newton$ 
in Figure~\ref{fig_xray_profile}. The large PSF of the instrument
certainly smooths the aforementioned feature in the X-ray profile.
However, the profile still shows a $2.4~\sigma$ bump at the same
location.

\section{DISCUSSION}

\subsection{Discrepant Mass Profiles in the Literature}
The galaxy cluster Cl~0024+17 has been known for its large (a factor of two or
more)  mass discrepancy between gravitational lensing and X-ray results.
However, the  large mass discrepancy from different {\it lensing} analyses has not received
its due attention. We compare five different mass profiles found in the
literature in Figure~\ref{fig_various_mass_profiles}.  Although some of the
large discrepancy at small radii ($\lesssim20\arcsec$) is due to the difference
in the choice of the cluster center,  it is remarkable that different profiles
give a very wide range of mass density outside the location of the critical
curve ($\sim30\arcsec$).

Some of the discrepancy might be explained by mass-sheet degeneracy
in gravitational lensing. Mass-sheet degeneracy refers to the invariant lensing
observable under the following transformation of the mass density: $\kappa
\rightarrow \lambda \kappa + (1-\lambda)$. When $\kappa$ is small ($<<1$),
the transformation can be viewed as adding an additional sheet of mass
(hence, the term "mass-sheet "). When $\kappa$ is not small, the
transformation becomes more conspicuous in change of slope. Indeed,
most of the different mass profiles in Figure~\ref{fig_various_mass_profiles}
roughly overlap under this transformation with a proper choice of $\lambda$ and a centroid.

However, the J07 profile is unique in the sense that the density does not
decrease at $r\gtrsim50\arcsec$. As repeatedly discussed above, the mass
density increases from this point on until it drops at $r\sim75\arcsec$.
Therefore, it would be incorrect to claim that the peculiarity of the J07 mass profile
is attributed to unresolved mass-sheet degeneracy because the invariant transformation
does not change the sign of the slope!

If the multiple image identification by Zitrin et al. (2009)
is flawless and the bias in their photometric redshift estimation is
negligible, their mass profile should be regarded
as the most accurately calibrated profile to date because the
33 multiply-imaged sources at different redshifts certainly provide  strong constraints
on the mass slope. Nevertheless, it is important to remember that 
these strong-lensing features are not useful in constraining the mass profile 
at $r\gtrsim50\arcsec$, where no multiple images have been identified.
Hence, it may be possible that the J07 mass profile would be transformed to
match the Zitrin et al. (2009) profile at $r\lesssim50\arcsec$.
Nevertheless, the result should still  reveal the peculiar structure at $r\gtrsim50\arcsec$ (with
reduced density because $\lambda$ must be greater than unity in this case).
A further analysis combining the strong lensing constraints of Zitrin et al. (2009) with the
weak-lensing data is a subject of our future investigation (M. Jee et al. in
prep).

The diverse results in the literature shown in Figure~\ref{fig_various_mass_profiles} clearly
demonstrates that the mass structure of the cluster Cl~0024+17 is still disputed
in the community, and thus more concerted efforts toward a better mass model, such as 
spectroscopic redshift survey of the multiply-lensed sources, high-resolution infrared imaging aimed for 
better photometric redshift estimation of background galaxies, etc., are desired. On the
other hand, it is important to investigate the issue with different observables. Our
study of the intracluster stars as potential tracers of the underlying dark matter structure
is one such approach, and here we argue that the peculiar features of the ICL that we 
report in this paper strongly favors the J07 result among the five
models. Of course, it is still an open question how well the intracluster stars trace
the underlying dark matter. While there is a consensus that intracluster stars
are not bound to individual galaxies but to the cluster potential, the answer
to this question certainly depends on when, where, and how the ICL is produced.
We discuss the issue in \textsection\ref{section_is_icl_dm_tracer}.

\subsection{Interpretation of the Similarity between ICL and Mass Profiles
\label{section_is_icl_dm_tracer}}

If intracluster stars are dominantly produced by truncation of galaxy halos during the
initial formation of a cluster (e.g., Merritt 1984), the dynamical history of the
intracluster stars is as old as the cluster itself. Hence, in this scenario we expect
to observe the strongest correlation between ICL and mass distributions.
Many $N$-body simulations, however, indicate that other mechanisms such as 
high-speed encounters between cluster galaxies (Moore et al. 1996), stripping from a cluster potential (Byrd \& Valtonen 1990), 
stripping in galaxy groups falling into the cluster (Mihos 2004), etc., 
might also contribute significantly to the overall production, which then would
make it less straightforward to interpret the ICL structure in conjunction with the mass distribution.
It is an open question which of these mechanisms is most dominant.

Recently, Murante et al. (2007) investigated the origin of the intracluster stars through
a cosmological hydrodynamical simulation. They reported that a dominant fraction ($\sim50$\%)
of the intracluster stars are produced in the formation of the most massive galaxy at the cluster
center, which is consistent with the observation that
the ICL is more centrally concentrated than
the cluster galaxies (e.g., Zibetti et al. 2005).
If cluster galaxies are poissonian, albeit sparse, tracers of underlying dark matter, this suggests
that perhaps the ICL profile might be more peaked also than the cluster mass profile.

With this view in mind, the resemblance of the ICL profile to the mass profile shown in
Figure~\ref{fig_masstoicl} may seem somewhat surprising. However, it is important to remember that
the ICL profiles are rescaled to match the mass profile, and thus it is trivially obvious that 
the agreement in the range $20\arcsec \lesssim r \lesssim 50\arcsec$ indicates nothing more than that both
the ICL and mass profiles monotonously decreases in this regime. What deserves our attention,
nonetheless, is that both profiles change the slope at $r\sim50\arcsec$  by the same amount (of
course, after rescaling is applied), giving a good agreement out to $r\sim70\arcsec$. 
The average ICL levels between the $r\sim20\arcsec$ and $r\sim50\arcsec$ regions differ
by a factor of five while in mass the density ratio is less than a factor of two.
Because the mean slope at $50\lesssim r \lesssim70\arcsec$ is much smaller than the value at $r\lesssim50\arcsec$, 
the ICL profile, although being much steeper than the mass profile, could be brought to agree after rescaling.

Therefore, our results presented here should not be interpreted as a demonstration that the
intracluster stars in Cl~0024+17 can be used as direct (i.e., poissonian) tracers of the underlying dark
matter. Our ICL profile measurement supports the view that the dominant ICL production mechanism
is associated with the formation of the brightest cluster galaxies at the cluster center, and
thus their distribution is expected to be more centrally concentrated on average. However, dynamically old
intracluster stars are given the chance to travel further away from the cluster center and
mix with the cluster halo dark matter. Hence,  the ICL profile should represent 
the dynamical history of the cluster in the central region, as well as the ICL
production history.

The expectation that cluster outskirts might be severely depleted of intracluster stars may aid us to
explain the difference between the mass and ICL profiles at $r\gtrsim72\arcsec$. As already 
discussed in \textsection\ref{section_comparison}, the ICL profiles change their slopes
more suddenly at $r\sim72\arcsec$ than the mass profile, and hereupon the ICL levels quickly approach
the background level. It is plausible that this is because the initial ICL level is already low
in this region before the dark matter structure is disrupted. Alternatively, we can
consider the possibility that the J07 mass profile is biased at $r\gtrsim75\arcsec$, and
the real mass profile is similar to the level that is indicated by the ICL profiles.
The radial mass profile in J07 does not represent the complete azimuthal average
at $r\gtrsim85\arcsec$, and thus the employed regularization can progressively bias
the mass density high as $r$ increases.

\subsection{Is the Cluster Galaxy Distribution Inconsistent with the Mass Profile
of J07?}

The current, well-accepted, structure formation paradigm postulates that galaxies are biased tracers of
hosting dark matter. Gravitational lensing studies of galaxy clusters have demonstrated that even in
clusters of complex morphology the spatial correlation between galaxies and dark matter is strong 
(e.g., Hoekstra et al. 2000; Jee et al. 2005; Clowe et al. 2006), although Mahdavi et al. (2007)
recently claims that a puzzling counter example is seen in Abell 520.

If galaxies and dark matter indeed follow each other, can we detect the ringlike structure of J07
in the cluster galaxy distribution? Of course, given that the ringlike structure of J07
has an only 5\% contrast, it is impossible to detect the feature as an enhancement of galaxy number density around the ``ring"
even if we improve the statistics by azimuthally averaging the numbers. Figure~\ref{fig_galaxy_vs_mass}
shows the projected number density of the cluster members as a function of radius
($0.37<z<0.41$) obtained from the
publicly available catalog of Moran et al. (2005), which extends the Czoske et
al. (2001) catalog.
The shot noise at $r\sim75\arcsec$ is $\sim45$\%, which is about 9 times higher than the contrast; we did not
correct for the completeness of the spectroscopic survey, which is negligible in this region compared with
the shot noise.

Apart from the issue of detecting the ringlike structure of J07, it is still interesting to
examine if the overall galaxy number density profile is consistent with the mass profile.
As mentioned by J07, the mass profile at $r>80\arcsec$ is measured from
an incomplete circle. Therefore, in order to enable a fair comparison, we should limit the fitting range 
to $r\leq80\arcsec$. The solid line in Figure~\ref{fig_galaxy_vs_mass} shows that
the J07 mass profile is consistent with the galaxy distribution ($\chi^2/d.o.f.=0.45$)
within $r\leq80\arcsec$. When we instead try the fit with the data points at $r\leq100\arcsec$, the result (dotted)
is severely affected by the data at $r\ge80\arcsec$, giving $\chi^2/d.o.f.=1.92$.
Qin et al. (2008) used the latter comparison to argue that the galaxy distribution is inconsistent with
the J07 mass profile. Because the mass profile of J07 at $r>80\arcsec$ does not represent the azimuthal average, 
we do not agree with their interpretation.
 
How many galaxies do we need to detect the peculiar dark matter structure? 
Within the $r=100\arcsec$ ($80\arcsec$) circle, there are 99 (84) $0.37<z<0.42$ galaxies in the
Moran et al. (2005) catalog. The mass density at the $r\sim75\arcsec$ bump is about 5\% higher than the value
at $r\sim50\arcsec$. If we want to detect this rising mass density in the $50\arcsec<r<75\arcsec$ region with the galaxy
distribution at the 2 $\sigma$ level, we need $\sim50$ times more galaxies, which is
even greater than the total number of objects ($\sim4000$) detected in the ACS image!
Our photo-z analysis estimates that about 350 out of 4000 galaxies are in the
redshift range $0.3 < z < 0.5$, and this number certainly does not help us to reveal the ringlike
dark matter structure.

\section{CONCLUSIONS}

Our main results of the investigation of the ICL and ICM in the galaxy cluster Cl~0024+17 are summarized as follows.

\begin{itemize}
\item The ICL profile closely resembles the peculiar dark matter structure reported in J07, which
stops decreasing at $r\sim50\arcsec$ ($\sim265$ kpc) and slowly increases until it turns over 
at $r\sim75\arcsec$ ($\sim397$ kpc).
\item The feature is present in both ACS and Subaru images in nearly all available passband images with a significance
in general higher in red filters.
\item The ICM profile shows a bump at $r\sim60\arcsec$ ($\sim318$ kpc). The radius of the ICM ring
      is $\sim20$\% smaller than that of the dark matter ring.
\item The two dimensional map of the diffuse light in the cluster shows that the ICL distribution
is elongated toward Northwest following cluster members.
\item The ICL color is consistent with the cluster red-sequence color at small radii ($r<40\arcsec$), but
      becomes bluer for increasing radii.
\item The ICL fraction of Cl~0024+17 is $\sim28$\% ($\sim37$\%) in F775W and F850LP (F475W, F555W, F625W) without
      any strong indication of radial dependence. Considering
      the redshift of the cluster, these values are likely to be at the high end of the distribution.      
\end{itemize}

These results not only support the presence of the peculiar dark matter structure
of J07
revealed by coherent fluctuation of the background galaxy shapes across the
$r\sim75\arcsec$ circle, but also
demonstrates the usefulness of the ICL as a powerful probe of dark matter substructure
in galaxy clusters. 

Theoretically, however, it is still disputable whether or not the ICL distribution
for a $z=0.4$ cluster should reveal the underlying dark matter.
Numerical simulations by Rudick et al. (2006) show that the ICL morphology is
a strong function of time. At early epoch most intracluster stars reside
in outer halos of individual galaxies while with time more tidal and
filamentary structures develop. As the evolution progresses further, it is shown
that these transitory features disappear into the more diffuse, common cluster halo.
If we associate the redshift $z=0.4$ with the young dynamical age in ICL evolution,
it is difficult to understand why the cluster's ICL follows the dark matter structure so remarkably well
as shown in this work. On the other hand, if Cl~0024+17 is regarded dynamically already mature
based on its optical richness or mass, the similarity between the ICL and dark matter structures
may not be surprising.
The high ICL fraction of the cluster supports the possibility that Cl~0024+17 is already
mature in the context of the ICL evolution. Alternatively, it is also plausible that the
high-speed collision 1-2 Gyrs ago might have liberated unusual amount of stars from the cluster galaxies.

The detection of the ICM bump at $r\sim60\arcsec$ opens an important opportunity to study the dynamical interplay
between the intracluster medium and the underlying dark matter. One simplistic
scenario that one can make out of this
feature is that both the ICM and dark matter ringlike structures have the same dynamical origin,
and both rings (or shells) are expanding. In this case, the expansion of the ICM ring must lag behind the dark matter ring
because of the ram pressure, which may explain the observed $\sim15\arcsec$ ($\sim80$ kpc) offset.

In a recent dark matter only simulation, ZuHone et al. (2008) claim that a bump can arise in cluster
collisions only when the initial particle velocity distribution is circular, and
the feature becomes a ``shoulder" as the tangential anisotropy decreases.
Our re-examination of the J07 collision simulation reveals
that the initial orbits of the particles in the cluster core were indeed
biased toward tangential anisotropy although the velocity anisotropy
is not as extreme (i.e., circular) as shown by ZuHone et al. (2008).
As the distinction between ``bump" and ``shoulder" is whether or not the slope of the post-collision
mass profile in the intermediate range (i.e.,$50\arcsec<r<75\arcsec$) is positive,
we believe that many other parameters such as the slope of the initial density profile,
the fraction of the gas particles, the relative mass ratio/speed of the two subclusters, etc.
should be also 
considered together with the initial
distribution of the particle orbits in order to fully address the issue.
Nevertheless, it is clear from the ZuHone et al. (2008) simulation that 
the creation of the ringlike structure, if the high-speed collision is to be the cause as suggested by J07,
requires some degree of tangential anisotropy at least in the core of the cluster. 

We acknowledge Maxim Markevitch for first informing us of the presence of the ringlike structure 
in the Chandra image
of Cl 0024+17. We would like to thank Bernard Fort for suggesting this study at the 23rd IAP colloquium.
Furthermore, we would like to thank Tony Tyson, Holland Ford, David Wittman, and John Blakeslee for useful discussions.
This research is in part supported 
by the grant from the TABASGO foundation 
awarded in the form of the Large Synoptic Survey Telescope Cosmology Fellowship.

\clearpage

\begin{figure}
\plotone{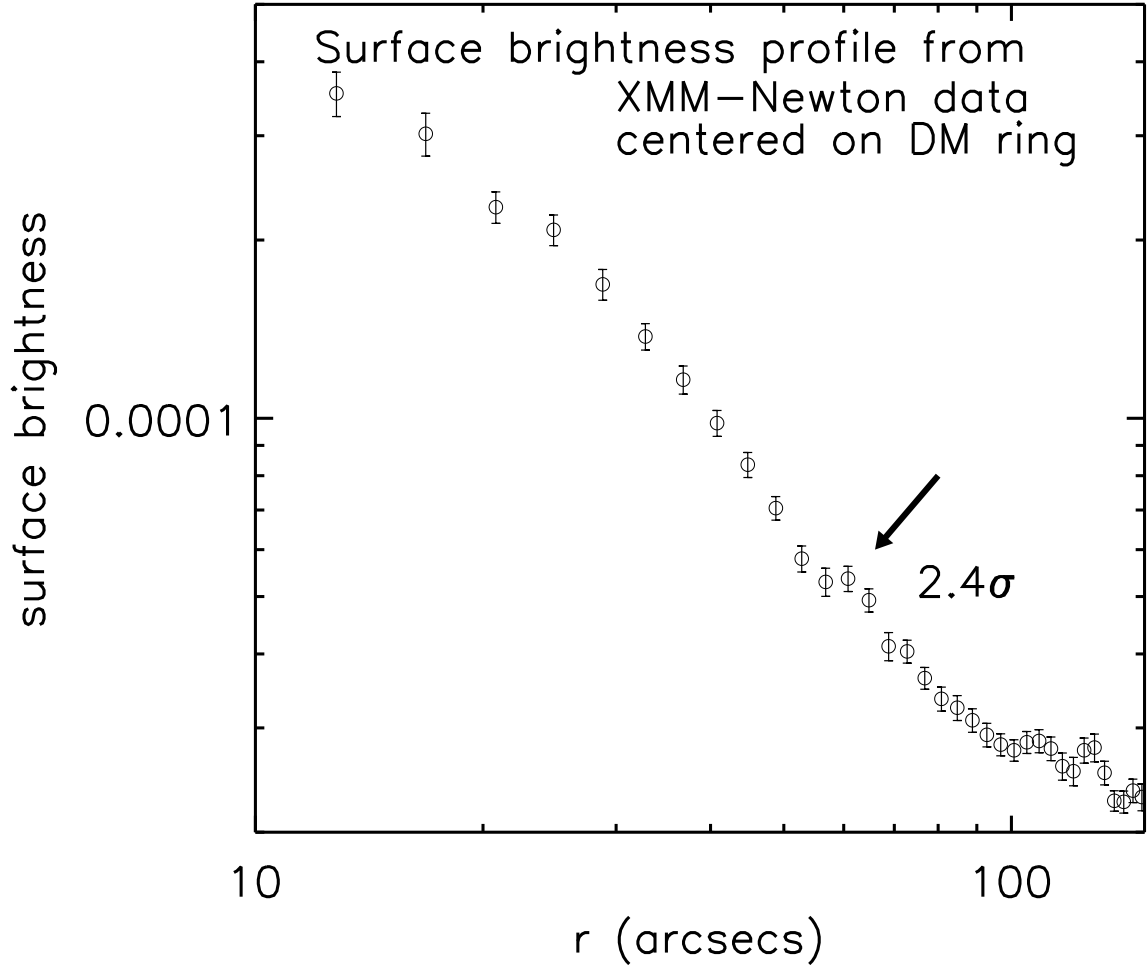}
\caption{X-ray surface brightness profile from the XMM-$Newton$ data. The bump at $r\sim60\arcsec$ seen 
in Figure~\ref{fig_xray_profile} is
somewhat smoothed by the larger PSF but still visible as a shoulder. 
The feature is difficult to observe if we chose the X-ray peak as the origin of the
radial bin as in the case of Figure~\ref{fig_xray_profile}. The scale is arbitrary.
The neighboring bins do not overlap.
\label{fig_xray_profile_xmmnewton}
}
\end{figure}

\begin{figure}
\plotone{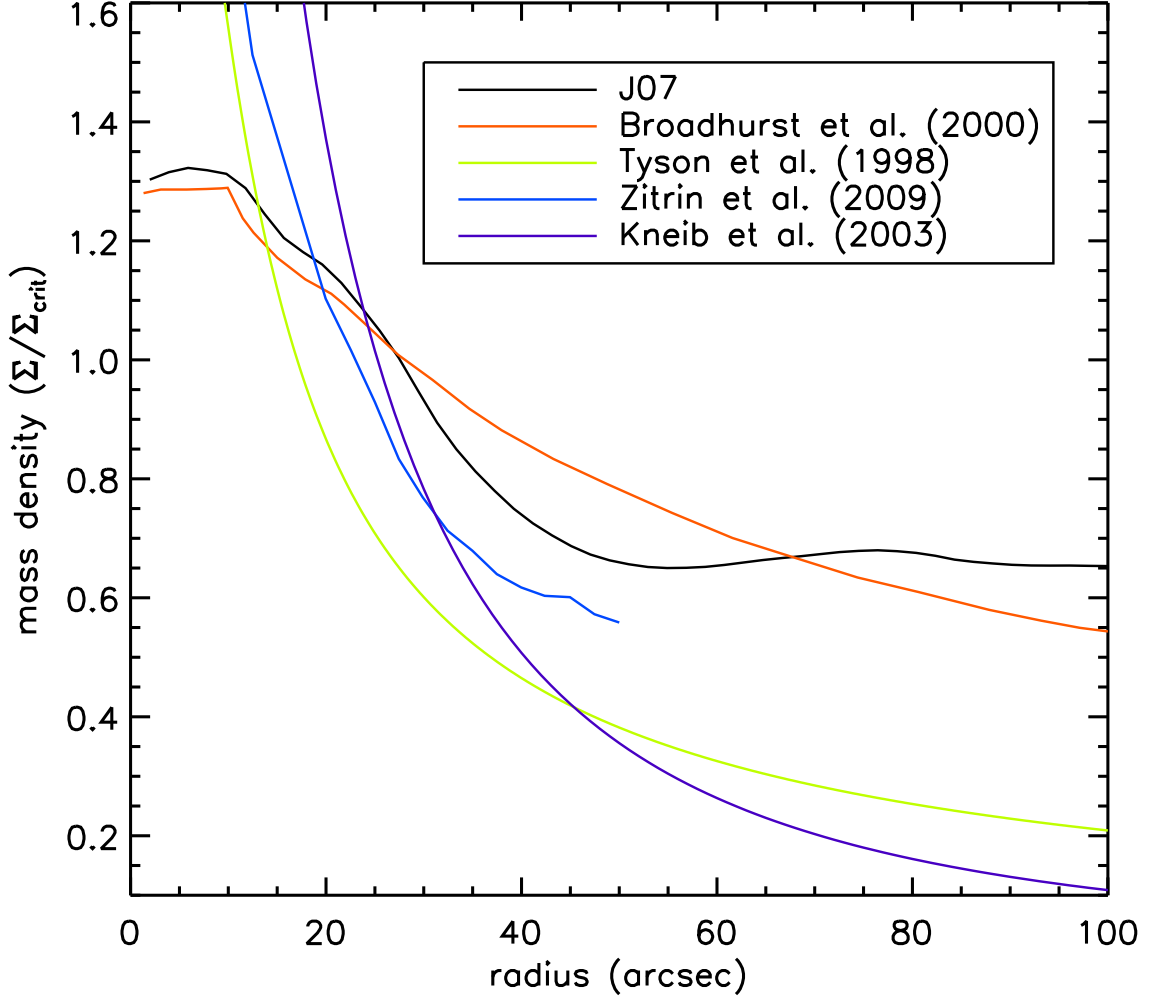}
\caption{Discrepant mass profiles of Cl~0024+17 in the literature.  The unit is
the critical lensing mass density at the $z=3$ reference redshift.  For the
Broadhurst et al. (2000) results, we reproduce their two-dimensional mass map, and 
measure the profile by placing the center at the geometric center of the
ringlike structure of J07.  For the Tyson et al. (2000) and Kneib et al. (2003)
we use their parametric fit to the mass profiles. The Zitrin et al. (2009)
profile shown here is their best-fitting model (read off from Figure 5 of the
paper).
\label{fig_various_mass_profiles}
}
\end{figure}

\begin{figure}
\plotone{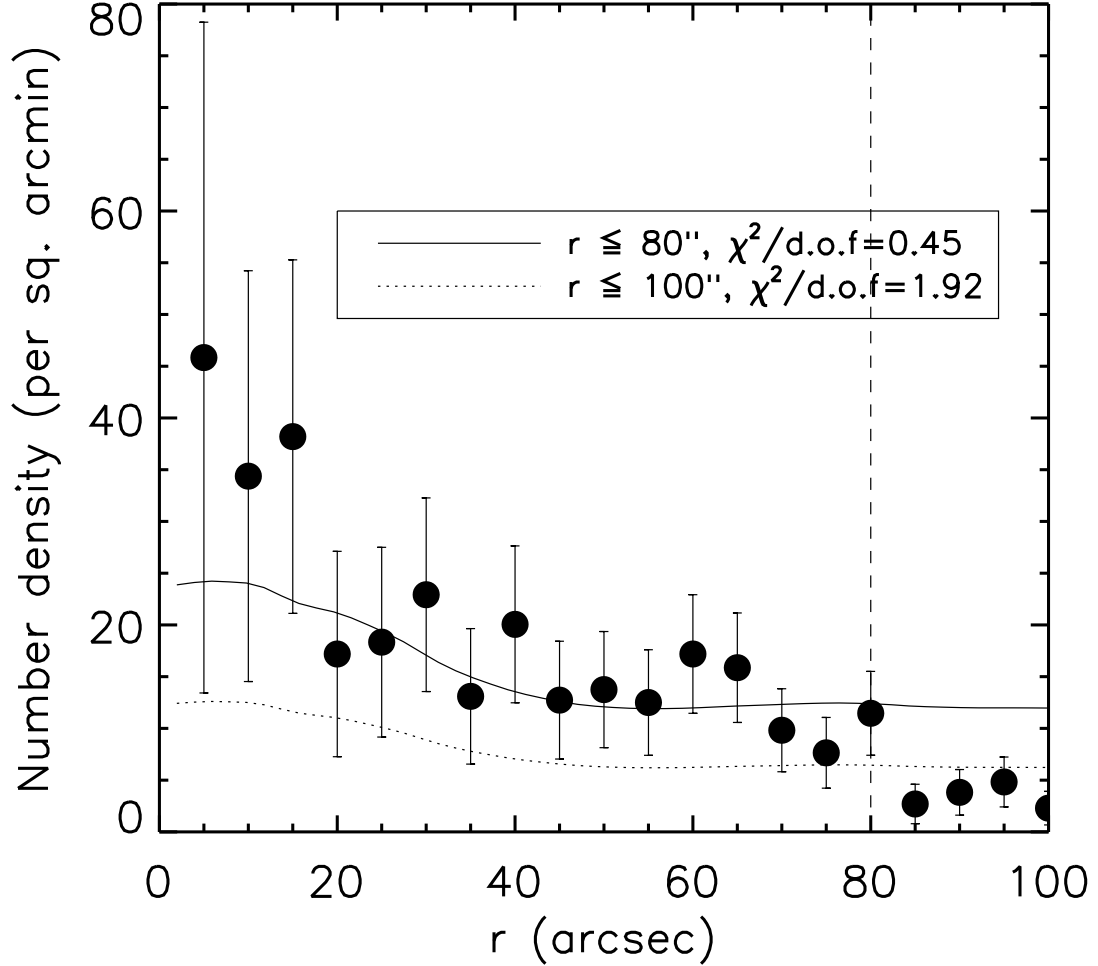}
\caption{Projected galaxy number density versus radius. We use the publicly available spectroscopic
catalog of Moran et al. (2005) to plot the number density of the spectroscopic members in the cluster
as a function of radius (filled circle). Because the mass profile of J07 is 
progressively biased beyond $r\gtrsim80\arcsec$ (dashed line) because of the limited field of view, 
a fair comparison may require limiting the range to fit at $r<80\arcsec$.
This gives a goodness-of-the-fit of $\chi^2/d.o.f=0.45$ (solid), which illustrates that the galaxy distribution
is statistically consistent with the J07 mass profile. If the data points at $r>80\arcsec$ are included, 
the goodness-of-the-fit degrades to $\chi^2/d.o.f=1.92$ (dotted). However, interpretation needs a caution
because the result is severely affected by the data points at $r>80\arcsec$, which have smaller error bars yet 
with lower completeness.
\label{fig_galaxy_vs_mass}
}
\end{figure}

\clearpage

\begin{deluxetable}{ccccc}
\tabletypesize{\scriptsize}

\tablecaption{ACS Image Sky Levels}

\tablenum{1}

\tablehead{\colhead{Filter} & \colhead {Visits} & \colhead{Total Exp. Time} & \colhead{Mean Sky Level} & 
      \colhead{Sky Level Variation Between Visits} \\
           \colhead{} &  \colhead{}      & \colhead{ (s) }     & \colhead{ (mag~$\mbox{arcsec}^{-2}$)} & \colhead{ (\%)} }
\tablewidth{0pt}

\startdata
F435W  & 5 & 6435  &  23.1 & 11\\
F475W  & 4 & 5072  &  22.9 & 4 \\ 
F555W  & 4 & 5072  &  22.6 & 3 \\ 
F625W  & 7 & 8971  &  22.2 & 8 \\ 
F775W  & 8 & 10144 &  22.2 & 2 \\ 
F850LP & 6 & 16328 &  22.1 & 2 \\ 
\enddata
\end{deluxetable}
                    
\begin{deluxetable}{ccccc}
\tabletypesize{\scriptsize}

\tablecaption{Subaru Image Sky Levels}

\tablenum{2}

\tablehead{\colhead{Filter} & \colhead{Visits} & \colhead{Total Exp. Time} & \colhead{Mean Sky Level at zenith} & 
      \colhead{Sky Level Variation Between Visits} \\
           \colhead{}  & \colhead{}     & \colhead{ (s) }     & \colhead{ (mag~$\mbox{arcsec}^{-2}$)} & \colhead{ (\%)}  }
\tablewidth{0pt}

\startdata
$B$           & 3  & 3,600  &  22.6 & 2  \\
$R_c$         & 11 & 5,280  &  21.2 & 3  \\ 
$z^{\prime}$  & 8  & 1,980  &  19.2 & 5  \\ 
$NB_{912}$      & 6  & 10,800 &  20.7 & 11 \\

\enddata
\end{deluxetable}

\begin{deluxetable}{ccccc}
\tabletypesize{\scriptsize}
\tablecaption{ACS Sky Flat Accuracy}
\tablenum{3}

\tablehead{\colhead{Filter} & \colhead{Number of Used Images} & \colhead{Deviation from Pipeline Flats} & 
      \colhead{Accuracy} \\
           \colhead{}  & \colhead{} & \colhead{ (\%)} & \colhead{ (\%)} }
\tablewidth{0pt}
\startdata
F435W  &90  &  0.5  &  0.08  \\
F475W  &138 &  0.4  &  0.07  \\
F555W  &92  &  0.4  &  0.06  \\ 
F625W  &155 &  0.5  &  0.07  \\
F775W  &154 &  0.4  &  0.07  \\
F850LP &312 &  0.7  &  0.08  \\
\enddata
\end{deluxetable}

\begin{deluxetable}{ccc}
\tabletypesize{\scriptsize}
\tablecaption{Subaru Sky Flat Accuracy}
\tablenum{4}
\tablehead{\colhead{Filter} & \colhead{Number of Used Images} & \colhead{Accuracy}   \\
           \colhead{}     & \colhead{  }     & \colhead{ (\%)} }
\tablewidth{0pt}
\startdata
$B$            &49  &   0.08 \\
$R_c$          &138 &   0.05 \\
$z^{\prime}$   &155 &   0.05 \\
$NB_{912}$     &41 &    0.09 \\
\enddata
\end{deluxetable}

\begin{deluxetable}{cccc}
\tabletypesize{\scriptsize}
\tablecaption{Errors in Subaru Background Level Estimation}
\tablenum{5}
\tablehead{\colhead{Filter} & \colhead{Fraction of Sky} & \colhead{Surface Brightness}  \\
           \colhead{}     & \colhead{ (\%)  }     & \colhead{ ($\mbox{mag}~\mbox{arcsec}^{-2}$) }}
\tablewidth{0pt}
\startdata
$B$          &    0.05 & 30.9\\
$R_c$        &    0.04 & 29.7\\
$z^{\prime}$ &    0.04 & 27.6\\
$NB_{912}$   &    0.08 & 28.4\\
\enddata
\end{deluxetable}

\begin{deluxetable}{cccc}
\tabletypesize{\scriptsize}
\tablecaption{ICL fraction in Cl 0024+17}
\tablenum{6}
\tablehead{\colhead{Filter} & \colhead{$r<200$ kpc} & \colhead{$r<500$ kpc}  \\
\colhead{} & \colhead{(\%)} & \colhead{(\%)} }
\tablewidth{0pt}
\startdata
F475W  &    $35\pm6$  &    $38\pm12$\\
F555W  &    $35\pm4$   &    $37\pm9$\\
F625W  &    $35\pm2$   &    $35\pm5$\\
F775W  &    $27\pm1$  &    $26\pm3$\\
F850LP  &   $29\pm1$   &    $30\pm3$\\
\enddata
\end{deluxetable}

\clearpage

\end{document}